\documentclass[pra,twocolumn]{revtex4}
\usepackage{hyperref}
\usepackage{amssymb, amsmath}
\usepackage{graphicx}
\usepackage{color}
\usepackage{empheq}
\newcommand{\bra}[1]{\left\langle {#1} \right|}
\newcommand{\ket}[1]{\left| {#1} \right\rangle}

\newcommand{\avg}[1]{\left\langle {#1} \right\rangle}
\newcommand{\bracket}[3]{\left\langle {#1} | {#2} | {#3} \right\rangle}
\newcommand{\bbracket}[3]{\bigl\langle {#1} \bigl| {#2} \bigr| {#3} \bigr\rangle}
\newcommand{\sbracket}[3]{\langle {#1} | {#2} | {#3} \rangle}

\newcommand{\beq}{\begin{equation}}
\newcommand{\eeq}{\end{equation}}
\newcommand{\bea}{\begin{eqnarray}}
\newcommand{\eea}{\end{eqnarray}}

\newcommand{\dbl}[1]{\bar{#1}}
\renewcommand{\k}{\textbf{k}}
\renewcommand{\l}{\textbf{l}}
\renewcommand{\a}{\textbf{a}}
\renewcommand{\b}{\textbf{b}}

\newcommand{\comment}[1]{}

\renewcommand{\k}{\textbf{k}}
\renewcommand{\l}{\textbf{l}}
\renewcommand{\a}{\textbf{a}}
\renewcommand{\b}{\textbf{b}}
\newcommand{\x}{\textbf{x}}

\newcommand{\theorem}[1]{{\it #1}}
\newcommand{\proof}{{\it Proof. }}

\begin{document}

\title{Quantum Noise of Free-Carrier Dispersion in Semiconductor Optical Cavities}

\author{Ryan Hamerly}\email{rhamerly@stanford.edu}
\author{Hideo Mabuchi}
\affiliation{Edward L.\ Ginzton Laboratory, Stanford University, Stanford, CA 94305}
	
\date{\today}
	
\begin{abstract}
	This paper derives Langevin equations for an optical cavity where the dominant nonlinearity arises from free-carrier dispersion.  We define a generalized Wigner function, compute a Fokker-Planck equation that approximates the master equation, and convert this to a system of stochastic differential equations.  These equations are similar to the Wigner equations for an optical Kerr cavity, but have additional noise terms due to the incoherent carrier excitation and decay processes.  We use these equations to simulate a phase-sensitive amplifier and latch and compare the results to a Kerr model.
\end{abstract}

\maketitle

Optical logic requires a platform that is fast, low-power and scalable to compete with electronics.  In the past decade, nano-photonics has advanced to the point where optical cavities of size $\lesssim (\lambda/n)^3$ and $Q$ factors $\gtrsim 10^4$ can be fabricated with standard techniques \cite{Notomi2010}.  The hope is that these cavities can be used to amplify the optical nonlinearity of materials or defects and perform all-optical logic for communications and computing at speeds and energy scales comparable to electronics.

Free-carrier dispersion is a promising nonlinearity for low-power optical logic.  The effect arises in all semiconductors.  In a semiconductor, there is a filled valence band and an empty conduction band, and when photons are absorbed, they excite electrons from the valence band to the conduction band.  Each absorption creates two free carriers -- an electron and a hole -- which evolve independently and decay on some timescale set by the material and its geometry.  The carriers provide feedback to the optical field by altering the absorption of the material (free-carrier absorption) or its refractive index (free-carrier dispersion).  On timescales long compared to the free-carrier lifetime, it acts as an effective optical nonlinearity and can be used to construct switches, amplifiers and other logic elements.

Accurate, semiclassical models for free-carrier effects already exist, and these are valid when the carrier and photon number are very large \cite{LundstromBook, Bennett1990}.  However, the real promise of free-carrier effects lies in their application to low-power photonic computing.  In some materials, free-carrier effects are strong enough that switching can be achieved with as few as 100 photons per cavity.  In this regime, quantum effects become important and place fundamental limits on device performance.  For example, quantum fluctuations in the photon number add noise to quantum amplifiers \cite{Caves1982} and lead to spontaneous switching in optical memories \cite{Kerckhoff2011, Mabuchi2011SS}.  This motivates the need to develop a quantum model for the free-carrier nonlinearity that works at low photon numbers, similar to the models that exist for cavity quantum electrodynamics (QED) and $\chi^{(3)}$ (Kerr) systems \cite{Kimble1998, Agrawal1979}.

In this paper, we derive a quantum-mechanical model for the free-carrier nonlinearity, following the standard open quantum systems formalism used for cavity QED, optical parametric oscillators (OPO's) and $\chi^{(3)}$ systems \cite{Gardiner1985}.  However, simulating even a single cavity in this model is not practical, since the large number of available carrier modes makes the full Hilbert space exponentially large.  Using a method based on the Wigner function \cite{Santori2014, Lugiato1978, Gardiner1988}, we can reduce the master equation to a set of c-number Langevin equations that are simple to simulate.  These equations bear resemblance to semiconductor laser rate equations and Bloch equations found in the literature \cite{Haug1969, AgrawalDutta, Lindberg1988}.  An adiabatic elimination reduces the model further, giving a set of stochastic differential equations (SDE's) for the field, electron number, and hole number in the cavity.  The deterministic part of these equations matches the classical models found in the previous literature, but the noise terms are new -- and have a quantum origin.

Section~\ref{sec:model} introduces the quantum model for the free-carrier cavity.  In Section~\ref{sec:wigner}, we introduce the Wigner formalism and apply it to this model, deriving a set of stochastic differential equations (SDE's) which we simplify by invoking a weak-doping, fast-dephasing limit.  (Extensions of this model, incorporating two-photon absorption and free-carrier absorption, are treated in Appendix~\ref{sec:related}).

The free carrier model we derive, summarized in Equations (\ref{eq:05-csde-1}-\ref{eq:05-csde-4}) and (\ref{eq:sde-sc1}-\ref{eq:sde-sc2}), resembles the equations of motion for the Kerr cavity derived in \cite{Santori2014}, but there are extra noise terms.  The steady-state behavior, and correspondence to the Kerr cavity, are treated in Section~\ref{sec:ss}.  Next, the free-carrier SDE's are applied to simulate two devices: a phase-sensitive amplifier in Section~\ref{sec:amp}, and an all-optical SR-latch in Section~\ref{sec:latch}.  For the amplifier, the free-carrier device does not show squeezing, whereas its Kerr analog does.  For the latch, the spontaneous switching rate is larger for the free-carrier device, and the discrepancy grows as the latch's bistable states become more widely separated.

\section{Quantum Model}
\label{sec:model}

\begin{figure}[tbp]
\begin{center}
\includegraphics[width=1.00\columnwidth]{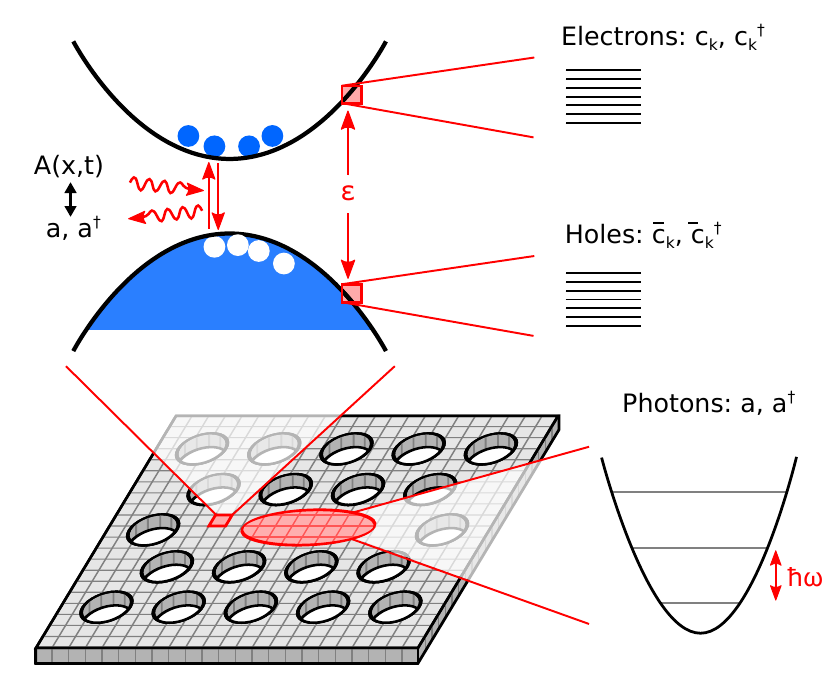}
\caption{(Color online) Electronic and optical modes of a semiconductor cavity.}
\label{fig:01-f2}
\end{center}
\end{figure}

Consider a single-mode optical cavity fabricated from an undoped semiconductor.  The optical degree of freedom can be represented as a harmonic oscillator, with the creation / annihilation operators $a, a^\dagger$.  In band theory, the electronic degree of freedom is represented by many uncoupled fermion modes.  For a two-band model, we have a single electron band and hole band.  Each mode has its own (fermionic) creation / annihilation operators -- $c_\k, c_\k^\dagger$ for electrons, $\bar{c}_\k, \bar{c}_\k^\dagger$ for holes, where $\k$ is the mode index.  The operator algebra is:

\begin{eqnarray}
	[a, a^\dagger] = 1,\ \ 
	\{c_\k, c_\l^\dagger\} = \{\bar{c}_\k, \bar{c}_\l^\dagger\} = \delta_{\k\l} \label{eq:05-opalg}
\end{eqnarray}

The operator algebra is key to the Wigner analysis in Section~\ref{sec:wigner}.  In short, one can define a generalized Wigner function for the quantum system if one can find a ``closed'' set of operators $\{X_i\}$, where the bath-averaged time derivatives $(dX_i/dt)_{\rm ad}$, $(d(X_iX_j)/dt)_{\rm ad}$, defined in Section \ref{sec:wigner}, are always functions of the $\{X_i\}$.  However, Wigner functions for fermionic operators require the use of Grassmann variables \cite{Cahill1999}, for which the analogy to classical phase space is less intuitive.  Thus, we identify fermion pairs and perform the following bosonization:

\begin{eqnarray}
	\sigma_{-\k} & = & c_\k \bar{c}_\k \label{eq:05-bos1} \\
	\sigma_{+\k} & = & \bar{c}_\k^\dagger c_\k^\dagger \label{eq:05-bos2} \\
	n_\k & = & c_\k^\dagger c_\k \label{eq:05-bos3} \\
	\bar{n}_\k & = & \bar{c}_\k^\dagger \bar{c}_\k \label{eq:05-bos4} \\
	Q_\k & = & n_\k\bar{n}_\k = \sigma_{+\k}\sigma_{-\k} = c_\k^\dagger\bar{c}_\k^\dagger \bar{c}_\k c_\k \label{eq:05-bos5}
\end{eqnarray}

This is similar to the operator algebra in an ensemble of two-level atoms, but there are some extra terms.  Analogous to the atom ensemble, the electronic polarization is given by $\sigma_{-\k}$.  However, the free-carrier system contains two number operators $n_{\k}$, $\bar{n}_{\k}$ rather than one, as well as a pairing operator $Q_{\k}$.  These arise because the electrons and holes in the free-carrier system have more freedom of movement: in an ensemble of atoms, each electron is confined to its parent atom and $n_\k = \bar{n}_\k = Q_\k$, while in a semiconductor these three quantities are no longer equal, since electrons and holes freely scatter between modes $\k$.  Operators (\ref{eq:05-bos1}-\ref{eq:05-bos5}) are bosonic because they are products of an even number of fermionic operators.  Note that several bosonic operators, namely $c_\k \bar{c}_\k^\dagger$ and $\bar{c}_\k c_k^\dagger$, are not included in (\ref{eq:05-bos1}-\ref{eq:05-bos2}) -- this is because we are interested in systems that respect charge conservation, while $c_\k \bar{c}_\k^\dagger$ and $\bar{c}_\k c_\k^\dagger$ violate it.

The bosonized operators are closed under commutation, with the following nonzero commutators:

\begin{eqnarray}
	{[}\sigma_{\pm \k}, n_\l] & = & \mp\delta_{\k\l}\sigma_{\pm \k} \\
	{[}\sigma_{\pm \k}, \bar{n}_\l] & = & \mp\delta_{\k\l}\sigma_{\pm \k} \\
	{[}\sigma_{\pm \k}, \sigma_{\mp \l}] & = & \pm\delta_{\k\l} (n_\k+\bar{n}_\k-1) \\
	{[}\sigma_{\pm \k}, Q_\l] & = & \mp \delta_{\k\l} \sigma_{\pm \k}
\end{eqnarray}
Note that these are all commutators, rather than anti-commutators, because the operators have been bosonized.

\subsection{Hamiltonian}

The Hamiltonian consists an an optical part $H_{\rm ph}$ which resembles a harmonic oscillator, an electronic part $H_{\rm el}$ given by the electronic band structure, and an interaction part $H_{\rm int}$ due to the $A\cdot p$ light-matter interaction.  It can be written as:

\beq
	H = \underbrace{\vphantom{\frac{n_k}{2}}\Delta_c a^\dagger a}_{H_{\rm ph}} + \sum_\k \Bigl[\underbrace{\Delta_\k \frac{n_\k + \bar{n}_\k}{2}}_{H_{\rm el}} + \underbrace{\vphantom{\frac{n_k}{2}}i g_\k (a^\dagger \sigma_{-\k} - a \sigma_{+\k})}_{H_{\rm int}}\Bigr]
	\label{eq:05-ham}
\eeq
where $\Delta_c = \hbar(\omega_c - \omega_{\rm ph})$ is the cavity resonance detuning, $\Delta_\k = E_{\k,c} - E_{\k,v} - \hbar\omega_{ph}$ is the detuning of the transition, and $g_\k$ is the atom-photon coupling.  The coupling can be expressed in terms of material parameters as shown in Appendix \ref{sec:materials}.

Hamiltonian (\ref{eq:05-ham}) resembles the cavity QED Hamiltonian.  This is because both systems contain an optical term, and electronic term, and a light-matter interaction of the $A\cdot p$ form.  Thus, it should not be surprising if free-carrier cavities exhibit many of the same phenomena observed in cavity QED, e.g.\ bistability, amplification, limit cycles \cite{Kwon2013}.

\subsection{External Interactions}
\label{sec:05-l}

In the cavity, the optical field is relatively well isolated from its environment.  The two primary interactions are optical absorption, which gives rise to particle-hole pairs and is treated through Eq.~(\ref{eq:05-ham}), and coupling to the external waveguide.  Because these couplings are usually quite weak, the optical field tends to retain its coherence in spite of them.

The same is not true for the carriers.  Many forces act to dephase, thermalize, and scatter the free carriers on very quick timescales (typically around $\sim 10-100$ fs) \cite{LundstromBook, Sabbah2002, Lin1987}.  Even for very poor cavities with $Q \lesssim 1000$, this is much faster than the photon lifetime.  The practical upshot of this will be that, on optical timescales, the ``coherent'' part to the carrier fields $\sigma_{\pm\k}$ can be adiabatically eliminated and only the ``slowly-varying'' carrier numbers $n_\k, \bar{n}_\k, Q_\k$ remain relevant to the system.

\begin{figure}[tbp]
\begin{center}
\includegraphics[width=1.00\columnwidth]{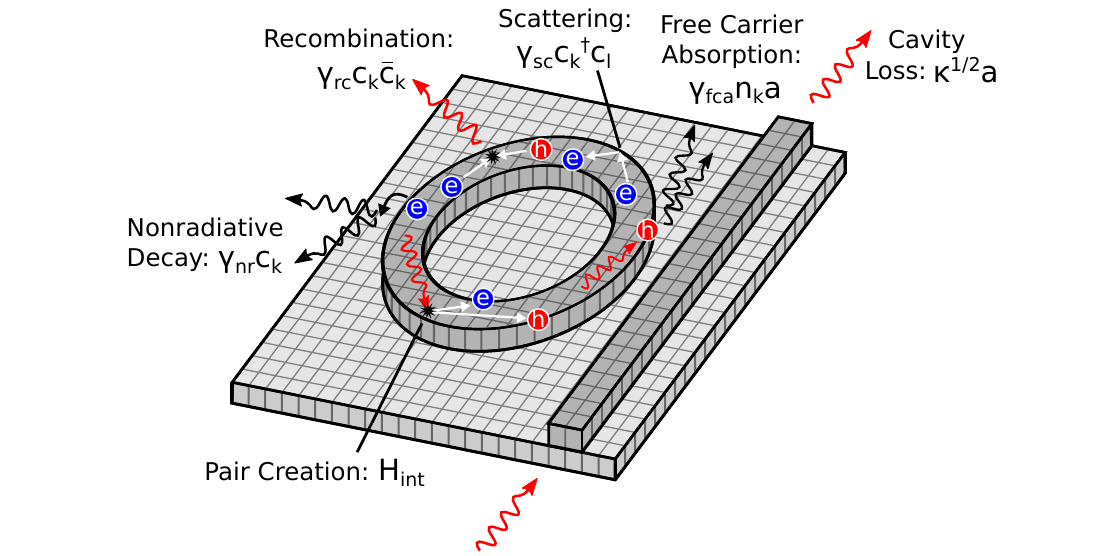}
\caption{(Color online) Major free-carrier effects in an optical cavity}
\label{fig:05-f3}
\end{center}
\end{figure}

For a bosonic, Markovian bath, external interactions can be treated by adding extra Lindblad terms to the Master equation \cite{Gardiner1985}.  The main external processes are given in Figure \ref{fig:05-f3} above.  As Lindblad terms, they are:

\begin{itemize}
	\item Cavity Loss, mediated by
		\beq
			L_{\rm cav} = \sqrt{\kappa}\;a
		\eeq
	\item Recombination, mediated by
		\beq
			L_{\rm rc,\k} =
				\sqrt{\gamma_{rc,\k}}\;\sigma_{-\k}
		\eeq
	\item Nonradiative Decay, mediated by
		\beq
			L_{{\rm nr},{\k}} =
				\sqrt{\gamma_{nr,\k}}\;c_{\k},\ \
				\sqrt{\bar{\gamma}_{nr,\k}}\;\bar{c}_{\k}
		\eeq
	\item Scattering / Dephasing, mediated by
		\beq
			L_{\rm sc,\k\rightarrow \l} =
				\sqrt{\gamma_{\k\rightarrow\l}}\,c_{\l}^\dagger c_{\k},\ \
				\sqrt{\bar{\gamma}_{\k\rightarrow\l}}\,\bar{c}_{\l}^\dagger \bar{c}_{\k}
		\eeq
\end{itemize}

Some materials also have significant free-carrier absorption and two-photon absorption.  For simplicity, these are not treated presently, but are discussed in Appendix~\ref{sec:otherprocess}.

\subsection{Single-Mode Theory}
\label{sec:singlemode}

To make the computation more tractable, we assume for now that all carrier modes $\k$ are identical.  This is not always a realistic assumption, and in Appendix \ref{sec:manymodes} we extend the result to non-identical modes.  But assuming identical modes for now, $\Delta_{\k}, g_{\k},$ and all the $\gamma_{\k}$'s become independent of $\k$.  Suppose that there are $N$ modes.  One can now define mode-summed operators:

\beq
\sigma_{\pm} = \sum_{\k} \sigma_{\pm \k},\ \mbox{etc.} \label{eq:sumop}
\eeq
This reduces the dimensionality of the state space from $5N+2$ to $7$.  In terms of these, the Hamiltonian and interaction terms are:

\bea
	\!\!\!\!H & \!=\! & \Delta_c a^\dagger a + \frac{1}{2}\Delta_e (n + \bar{n}) + i g (a^\dagger \sigma_{-} \!- a \sigma_{+}) \label{eq:slh-1} \\
	\!\!\!\!L_{\rm cav} & \!=\! & \sqrt{\kappa}\;a \\
	\!\!\!\!L_{\rm rc,\k} & \!=\! &
				\sqrt{\gamma_{rc}}\;\sigma_{-\k} \\
	\!\!\!\!L_{{\rm nr},\k} & \!=\! &
				\sqrt{\gamma_{nr}}\;c_{\k},\ \
				\sqrt{\bar{\gamma}_{nr}}\;\bar{c}_{\k} \\
	\!\!\!\!L_{\rm sc,\k\rightarrow \l} & \!=\! &
				\sqrt{\frac{\gamma_{\rm sc}}{2}}\;c_{\l}^\dagger c_{\k},\ \
				\sqrt{\frac{\bar{\gamma}_{\rm sc}}{2}}\;\bar{c}_{\l}^\dagger \bar{c}_{\k} \label{eq:slh-n}
\eea
Note that the sum operators used are bosonic, not fermionic.  The single-mode theory would not work if we had started with the fermionic operators.

\section{Wigner Function and SDE's}
\label{sec:wigner}

\subsection{Derivation from Quantum Model}

Under the quantum model described in Section \ref{sec:model} above, the state of the cavity is given by a density matrix $\rho$ and evolves according to the master equation:

\beq
	\frac{d\rho}{dt} = -i[H, \rho] + \frac{1}{2} \sum_k \left(2L_k \rho L_k^\dagger - L_k^\dagger L_k \rho - \rho L_k^\dagger L_k\right) \label{eq:master}
\eeq

Unfortunately, with an exponentially large Hilbert space, it is not practical to compute $\rho$ or its evolution.  To get around this problem, we express $\rho$ in terms of a generalized Wigner function, compute the equations and show that they can be approximated by a Fokker-Planck equation, and solve the Fokker-Planck equation stochastically using SDE's.

This approach was derived by Carter \cite{Carter1995} for optical fibers with a $\chi^{(3)}$ interaction; the same approach can be applied to optical cavities or cavity-based photonic circuits \cite{Santori2014}.  In both cases, there is an injective linear mapping between optical states $\rho$ and Wigner functions $W(\alpha, \alpha^*)$.  Gronchi and Lugiato~\cite{Lugiato1978} extended this method to weakly-coupled many-atom cavity QED.  In this case, in addition to an optical degree of freedom, one must also keep track of $N$ two-level atoms.  The procedure is to define a closed Lie algebra $\{ X_i \}$ of both optical and atomic operators, and a semiclassical phase-space that with c-number variables $\{ x_i \}$.  The generalized Wigner function is defined analogous to the optical function -- the Fourier transform of a characteristic function:

\beq
	W(x,t) = \int{d^ny\,e^{-i\sum_k x_k y_k} {\rm Tr}\left[e^{i\sum_k X_k y_k}\rho(t)\right]}
\eeq
In our case, the operator algebra consists of the optical and single-mode carrier operators (\ref{eq:sumop}) and is given as follows:

\beq
	X = \left[a, a^\dagger, \sigma_{-}, \sigma_{+}, n, \bar{n}, Q\right]
\eeq

This is a $7$-dimensional, operator-valued vector.  The Wigner function thus lives is a $7$-dimensional phase space, defined over the c-number variables:

\beq
	x = \left[\alpha, \alpha^*, v, v^*, m, \bar{m}, q\right]
\eeq

Under certain closedness conditions discussed in Appendix \ref{sec:closedness}, the Wigner function satisfies a generalized Fokker-Planck equation.  Truncating at second order, this reduces to a Fokker-Planck equation.  The validity of this truncation generally requires that nonlinear coupling constants be weak \cite{Santori2014}, and for two-level systems, that the number of atoms (carrier modes in this case) be large \cite{Lugiato1978}.  Both assumptions are true here.  As a solution to a Fokker-Planck equation, the Wigner function can be sampled stochastically by solving the following SDE's:

\beq
	dx_m = \mu_m\,dt + \sum_n R_{mn}\,dw_n \label{eq:fpe-sde}
\eeq
with $dw_n$ a Wiener process and

\bea
	\mu_m & = & C_m^{(1)}(x) \equiv \left(\frac{dX_m}{dt}\right)_p \label{eq:cum1} \\
	(RR^T)_{mn} & = & C_{mn}^{(2)}(x) \equiv \frac{1}{2}\left(\frac{d(X_mX_n-X_nX_m)}{dt}\right)_p \nonumber \\
	& & - x_m \left(\frac{dX_n}{dt}\right)_p - \left(\frac{dX_m}{dt}\right)_p x_n \label{eq:cum2}
\eea
where the time-derivatives are computed using the adjoint to (\ref{eq:master})

\beq
	\left.\frac{dA}{dt}\right|_{\rm ad} \equiv -i[A, H] + \frac{1}{2}\sum_k{2L_k^\dagger A L_k - L_k^\dagger L_k A - A L_k^\dagger L_k} \label{eq:05-ad}
\eeq
and $(\cdots)_p$ is defined so that normally ordered products return simple c-number polynomials, e.g.\ $(a)_p = \alpha$, $\frac{1}{2}(a^\dagger a + a a^\dagger)_p = \alpha^*\alpha$, $(a^\dagger a)_p = \alpha^*\alpha - \frac{1}{2}$, etc.

Computing the cumulant matrices $C^{(1)}$ and $C^{(2)}$ from the quantum model (\ref{eq:slh-1}-\ref{eq:slh-n}) is straightforward but very tedious, so we used {\it Mathematica} to derive the result.  The details are given in Appendix \ref{sec:fullsdes}; the SDE's are:

\begin{eqnarray}
	d\alpha & = & \left[\left(-\frac{\kappa}{2} - i\Delta_c \right)\alpha + g v\right]dt + d\xi_1 \label{eq:sde1} \\
	d\alpha^* & = & \left[\left(-\frac{\kappa}{2} + i\Delta_c\right)\alpha^* + g v^*\right]dt + d\xi_2 \\
	dv & = & \Bigl[-\alpha (N-m-\bar{m}) g \nonumber \\
	& & + \left(-\frac{\gamma_{tot}}{2} - i\Delta_e\right) v\Bigr]dt + d\xi_3 \\
	dv^* & = & \Bigl[-\alpha^* (N-m-\bar{m}) g \nonumber \\
	& & + \left(-\frac{\gamma_{tot}}{2} + i\Delta_e\right) v^*\Bigr]dt + d\xi_4 \\
	dm & = & \left[-g(\alpha v^* + v\alpha^*) - \gamma_{nr} m - \gamma_{rc} q\right]dt + d\xi_5 \\
	d\bar{m} & = & \left[-g(\alpha v^* + v\alpha^*) - \bar{\gamma}_{nr} \bar{m} - \gamma_{rc}q\right]dt + d\xi_6 \\
	dq & = & -\gamma_{sc} (q - m\bar{m}/N)dt + d\xi_7 \label{eq:sdeN}
\end{eqnarray}
where $\gamma_{tot} = \gamma_{sc} + \gamma_{rc} + \gamma_{nr} + \bar{\gamma}_{nr}$ and the noise processes $d\xi_i$ have the covariance matrix:

\beq
	d\xi d\xi^T = C^{(2)}\,dt
\eeq
where $C_{\rm cav}^{(2)} + C_{\rm int}^{(2)} + C_{\rm rc}^{(2)} + C_{\rm nr}^{(2)} + C_{\rm sc}^{(2)}$ is the sum of the terms in Eqs.~(\ref{eq:app1}-\ref{eq:app2}).  Note how Equations (\ref{eq:sde1}-\ref{eq:sdeN}) resemble both the Maxwell-Bloch equations and the Langevin equations for many-atom cavity QED derived by Gronchi and Lugiato~\cite{Lugiato1978}.  However, because of scattering between carrier modes, we need to keep track of $m$, $\bar{m}$ and $q$ separately.

\subsection{Approximations}
\label{sec:approx}

\subsubsection{Fast Dephasing, Nondegenerate Excitation}

Three approximations make these equations more tractable: fast-dephasing, nondegenerate excitation and the single-carrier approximation.  {\it Fast dephasing} assumes that the scattering rate $\gamma_{sc}$ and detuning $\Delta_e$ are faster than any other timescale in the system, thus

\beq
	\gamma_{nr}, \gamma_{rc}, \gamma_{fc} \ll \gamma_{sc}, \Delta_e \label{eq:05-fastdeph}
\eeq

This is related to the relaxation-time approximation that holds for most semiconductors \cite{LundstromBook}.  In useful, optimized free-carrier devices, all of the carrier timescales -- $\gamma_{nr}, \gamma_{rc}, \gamma_{fc}$ -- are of order the photon lifetime.  To achieve strong carrier effects, we generally have cavities with $Q \gtrsim 1000$, giving a photon lifetime of $\tau_{ph} \gtrsim \mbox{ps}$.  Ultrafast studies show that inter-mode scattering takes place on times of order 10--100 fs \cite{Sabbah2002, Lin1987}, giving scattering rates at least 10--100 times faster than any other timescale in the system.

Fast dephasing leads to an adiabatic elimination of the dipole terms $(v, v^*)$ and the pair density $q$.  These variables will be replaced by their steady-state values, and a new set of SDE's are obtained for the reduced basis $(\alpha, \alpha^*, m, \bar{m})$.

{\it Nondegenerate excitation} assumes that the number of carriers is much less than the number of carrier modes -- in other words, the valence and conduction bands are far from being degenerately filled with electrons or holes.  This approximation is invoked by setting

\beq
	m, \bar{m}, q \ll N
\eeq

This simplifies the equations of motion by discarding effects like absorption saturation that are negligible for low-power, high-$Q$ optical logic devices.  The resulting equations of motion are:

\bea
	\!\!\!\!d\alpha \!& = & \!\left[-\frac{\kappa+\eta}{2} - i\bigl(\Delta_c + \delta_c m+ \bar{\delta}_c\bar{m}\bigr)\right]\alpha\,dt + d\xi_\alpha \label{eq:05-csde-1} \\
	\!\!\!\!dm \!& = & \!\left[\eta\,\alpha^*\alpha - \gamma_{nr}m - \gamma_{rc}m\bar{m}\right]dt + d\xi_m \label{eq:05-csde-3}  \\
	\!\!\!\!d\bar{m} \!& = & \!\left[\eta\,\alpha^*\alpha - \bar{\gamma}_{nr}\bar{m} - \gamma_{rc}m\bar{m}\right]dt + d\xi_{\bar{m}} \label{eq:05-csde-4}
\eea
with noise terms

\bea
	d\xi_\alpha & = & -\sqrt{\kappa}\,d\beta_{\rm in} - \sqrt{\eta}\,d\beta_\eta \\
	d\xi_m & = & 2\sqrt{\eta}\,\mbox{Re}[\alpha^* d\beta_\eta] + \sqrt{\gamma_{nr} m}\,dw_m \nonumber \\
	& & + \sqrt{\gamma_{rc} m\bar{m}}\,dw_{rc} \\
	d\xi_{\bar{m}} & = & 2\sqrt{\eta}\,\mbox{Re}[\alpha^* d\beta_\eta] + \sqrt{\gamma_{nr} \bar{m}}\,dw_{\bar{m}} \nonumber \\
	& & + \sqrt{\gamma_{rc} m\bar{m}}\,dw_{rc}
\eea
where the $d\beta$'s are complex vacuum Wiener processes, e.g.\ $d\beta_\eta^*d\beta_\eta = \frac{1}{2}dt$, and the $dw$'s are real Wiener processes, e.g. $dw_m^2 = dt$.

In the equations above, we rescaled $\gamma_{rc}$ and defined a bandfilling carrier-dependent detuning $\delta_c$ and linear absorption $\eta$:

\bea
	\delta_c = \bar{\delta}_c & = & \frac{g^2}{\Delta_e - \frac{1}{2}i \gamma_{sc}} \label{eq:delta-bf} \\
	\eta & = & 2N\mbox{Im}[\delta_c] \label{eq:eta-bf}
\eea

Since $\Delta_e = \omega_e - \omega$, this function has one pole (for $\omega$) in the lower half-plane, $\omega = \omega_e - \frac{1}{2}i\gamma_{sc}$.  As a result, its real and imaginary parts satisfy the Kramers-Kronig relations.  The carrier-dependent dispersion and absorption are given by the real and imaginary parts of $\delta_c$, respectively.  Around $1/\delta_c$ carriers are needed to shift the cavity resonance by one linewidth; since this quantity is much smaller than $N$ under the nondegenerate approximation, it follows that $\mbox{Im}[\delta_c] \ll \mbox{Re}[\delta_c]$.  For a pure bandfilling effect, we can generally neglect the imaginary part.

The $d\beta_{\rm in}$ in (\ref{eq:05-csde-1}-\ref{eq:05-csde-4}) is the vacuum noise of the input field and $d\beta_\eta$ is the noise due to linear absorption; each behaves as a vacuum Wiener process $d\beta^* d\beta = \frac{1}{2}dt$ \cite{Santori2014}.  The $dw_m$, $dw_{\bar{m}}$ and $dw_q$ are real-valued noises due to carrier loss and recombination, and go as $dw^2 = dt$.

The noise term for $\alpha$ is fairly standard for open quantum systems: a sum of two vacuum noises.  The noise terms for $m$ and $\bar{m}$ have Poisson statistics: for each process with rate $R\,dt$, there is a corresponding noise term $\sqrt{R}\,dw$.  Since carrier generation involves photon absorption, one should not be surprised by the Poisson noise on this signal.  Likewise, since the carrier number is quantized and carrier decay is a random process, there should also be Poisson noise on the decay terms.

\subsubsection{Single-Carrier Approximation}
\label{sec:sca}

In many cases, the equations (\ref{eq:05-csde-1}-\ref{eq:05-csde-4}) can be reduced further by positing that $m = \bar{m}$ and introducing an {\it effective carrier number} $N_c$ equal to this quantity.  For example, it will hold if only one of the carrier species is relevant (for instance in silicon, where $\delta_c \gg \bar{\delta}_c$ due to the plasma effect \cite{Bennett1990}), if the recombination process $\gamma_{rc}$ is dominant, or if the number of recombination sites is limited (much smaller than the number of carriers) so that electrons and holes tend to decay together.  If any of these cases hold true, equations (\ref{eq:05-csde-1}-\ref{eq:05-csde-4}) become:

\bea
	d\alpha & = & \left[-\frac{\kappa+\eta}{2} - i(\Delta_c + \delta_c N_c)\right]\alpha\,dt + d\xi_\alpha \label{eq:sde-sc1} \\
	dN_c & = & \left[\eta\,\alpha^*\alpha - \gamma_{nr}N_c - \gamma_{rc}N_c^2\right]dt + d\xi_N \label{eq:sde-sc2}
\eea
with noise terms

\bea
	d\xi_\alpha & = & -\sqrt{\kappa}\,d\beta_{\rm in} - \sqrt{\eta}\,d\beta_\eta \\
	d\xi_N & = & 2\sqrt{\eta}\,\mbox{Re}[\alpha^* d\beta_\eta] + \sqrt{\gamma_{nr} N_c}\,dw_{nr} \nonumber \\
	& & + \sqrt{\gamma_{rc} N_c^2}\,dw_{rc}
\eea

In the sections below, we consider the behavior of optical logic devices assuming the single-carrier approximation.  Note that this is not strictly necessary -- it is equally feasible to use Eqs.~(\ref{eq:05-csde-1}-\ref{eq:05-csde-4}) to obtain similar qualitative results.

\section{Steady-State Behavior}
\label{sec:ss}

\begin{figure}[btp]
\begin{center}
\includegraphics[width=1.00\columnwidth]{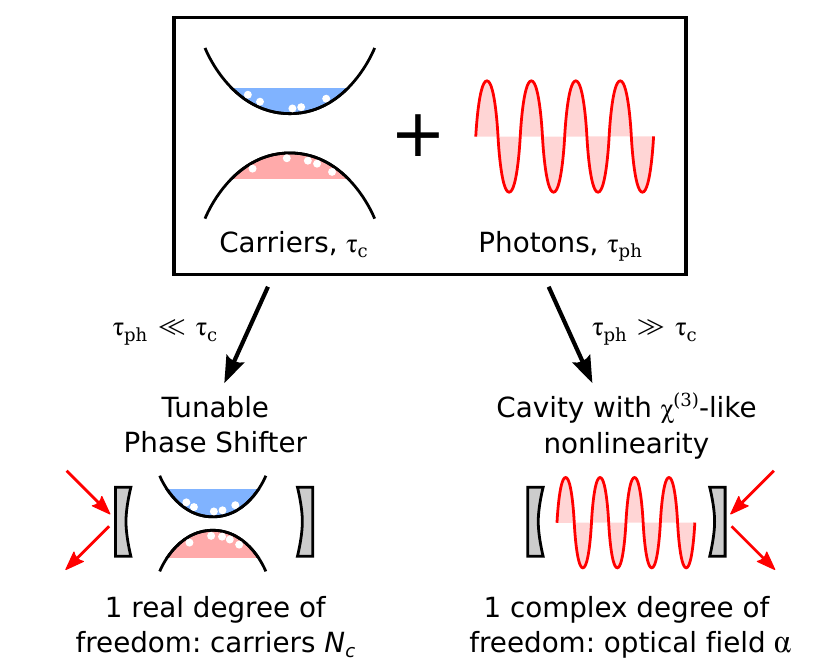}
\caption{(Color online) Adiabatic elimination of a free-carrier device into a tunable phase shifter (left) and a Kerr-like nonlinear cavity (right).}
\label{fig:06-f1}
\end{center}
\end{figure}

Consider now the case where (\ref{eq:sde-sc1}-\ref{eq:sde-sc2}) hold.  Suppose also that direct recombination is negligible ($\gamma_{rc} = 0$).  This is the limit to which III-V photonic crystals operated near the band edge, which have the best performance to date \cite{Nozaki2010}, belong (the parameters in Table \ref{tab:t1} are less than a factor of 5 from the state of the art).  In this limit, a free-carrier cavity modeled by equations (\ref{eq:sde-sc1}-\ref{eq:sde-sc2}) has two timescales -- an optical lifetime $\tau_{ph} \equiv 1/(\kappa+\eta)$ and a free-carrier lifetime $\tau_{c} \equiv 1/\gamma_{nr}$.

First, a steady-state limit is discussed.  This is the case when \textit{both} the carrier and photon lifetimes are much shorter than the relevant timescales.  Questions of thermal stability, for instance, can be treated in the steady-state limit.  Next the limiting case of $\tau_{ph} \gg \tau_c$, where the carrier population varies much faster than the photon population, is treated and we show that the free-carrier model reduces to a Kerr model with extra noise terms.

In this section, we work in normalized units by setting $k \equiv \kappa+\eta \rightarrow 1$.  Rates, time constants and powers are scaled by appropriate powers of $k$.  This allows our results to generalize to a wide range of systems spanning orders of magnitude in speed and size.

\subsection{Steady-State Limit}

In the steady-state case, we set all noise terms to zero and solve for $\dot{N}_c = \dot{\alpha} = 0$.  Solving for $\dot{\alpha} = 0$, the steady-state internal field $\bar{\alpha}$ can be related to $N_c$ and the input field $\beta_{in}$ as follows:

\beq
	\alpha = \frac{-\sqrt{\kappa} \beta_{in}}{\frac{\kappa+\eta}{2} + i(\Delta_c + \delta_c N_c)} \label{eq:06-ab}
\eeq

This is the familiar formula for the field in a resonant cavity, where the detuning $\Delta_c + \delta_c N_c$, depends on the free-carrier number.  Solving the $\dot{N}_c = 0$ equation gives $N_c = (\eta/\gamma_{rc})\alpha^*\alpha$.  This can be rearranged into a polynomial equation for $\alpha^*\alpha$:

\beq
	\kappa\,\beta_{in}^* \beta_{in} = (\alpha^*\alpha) \left[(k/2)^2 + \bigl(\Delta_c + (\eta\delta_c/\gamma_{nr}) (\alpha^*\alpha) \bigr)^2\right] \label{eq:06-cubic}
\eeq

\begin{table}[btp]
\centering
\begin{tabular}{c|l|c}
	\hline \hline
	Name & Description & Value (this paper) \\ \hline
	$k$ & Total linear loss $k = \kappa+\eta = \omega/Q$ & 0.5 ps$^{-1}$ \\
	$\kappa$ & Output coupling & $0.8k$ \\
	$\eta$ & Linear Absorption & $0.2k$ \\
	$\beta$ & 2-photon Absorption & 0 \\
	$\chi$ & Kerr & 0 \\
	$\delta_c$ & Free-Carrier Dispersion & $0.013k$ \\
	$\gamma_{rc}$ & Recombination & 0 \\
	$\gamma_{nr}$ & Free-Carrier Decay & $1.0k$ \\
	\hline\hline
\end{tabular}
\caption{System parameters used in the free-carrier simulations in this section.  Values given are similar to state of the art III-V photonic crystals.}
\label{tab:t1}
\end{table}

When the external power $P = \beta_{in}^*\beta_{in}$ is set, this is a cubic equation for the internal photon number $N_{ph} = \alpha^*\alpha$.  It is the same optical bistability cubic as the Kerr cavity \cite{Agrawal1979, Yurke2006}, with the effective Kerr nonlinearity:

\beq
	\chi_{\rm eff} = \frac{\eta\delta_c}{\gamma_{nr}}
\eeq

\begin{figure}[tbp]
\begin{center}
\includegraphics[width=1.00\columnwidth]{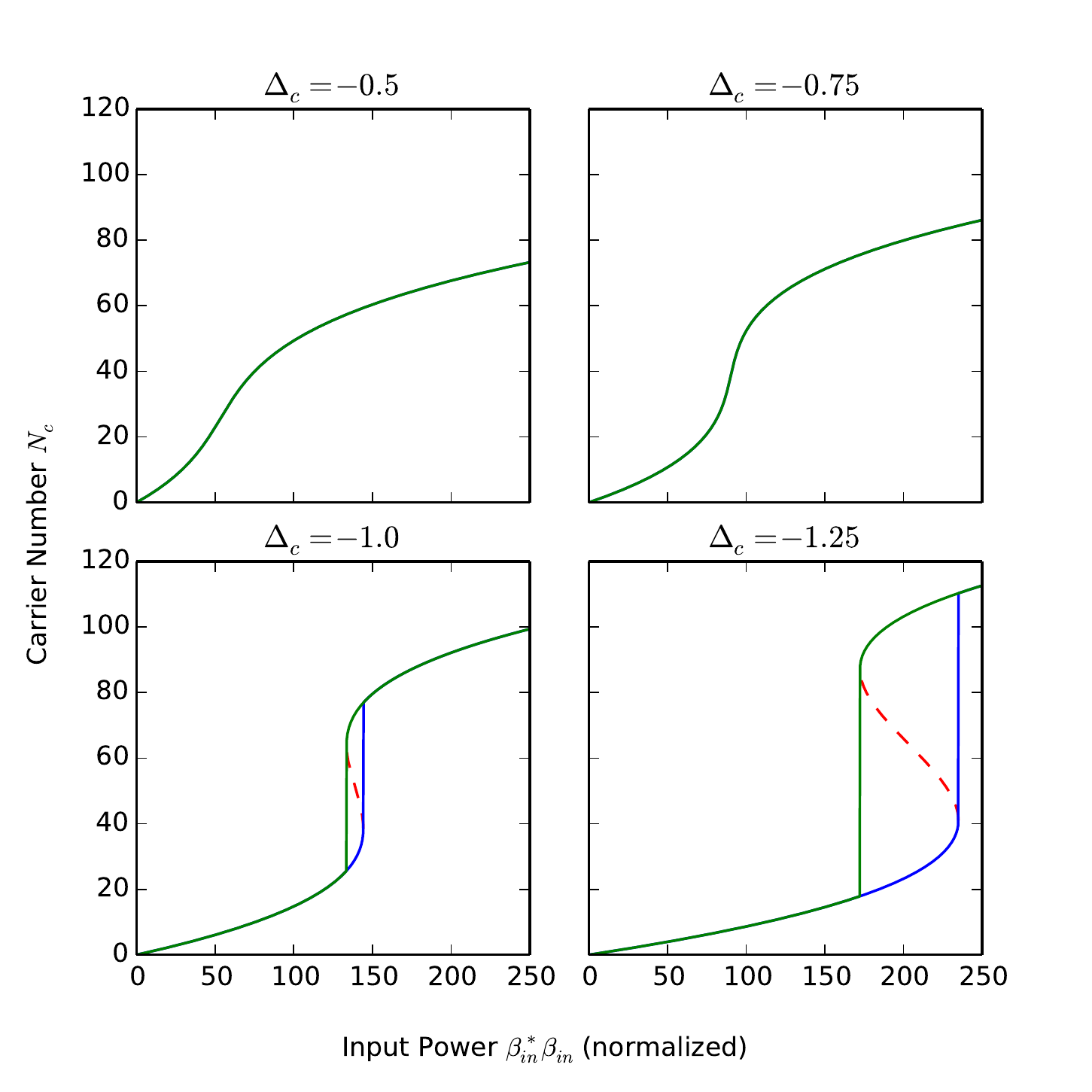}
\caption{(Color online) Steady-state solutions to $N_c$ for optical free-carrier cavity at different detunings.}
\label{fig:06-f2}
\end{center}
\end{figure}

One can solve the cubic (\ref{eq:06-cubic}) to obtain $\alpha^*\alpha$; it is not always uniquely defined.  Just like Kerr cavities and atom cavities, free-carrier cavities exhibit hysteresis and bistability, with both ``low'' and ``high'' intensity states being allowed for the same input power.  Figure \ref{fig:06-f2} shows the stable lower- and upper states, and an unstable middle-state, for varying values of $\Delta_c$.

The intuition behind this bistability is that, when the cavity is off resonance and a sufficiently large number of carriers are injected, it will shift back on resonance.  If there is a strong enough input, then a large power builds up inside the cavity and this large carrier population can be maintained through absorption, giving rise to the high state.  On the other hand, if there are no carriers to begin with, the cavity stays off resonance and there is never enough power in the cavity to raise the carrier number -- hence the low state.  Analytically, one can show that the bifurcation sets in when:

\beq
	\Delta_c < -\sqrt{\frac{3}{4}}\,(\kappa+\eta)
\eeq

Much of our intuition behind free-carrier nonlinearities comes from this steady-state picture.  It does not include any quantum effects or even any dynamics, but the shapes of the curves in Figure~\ref{fig:06-f2} suggest that the device could be used as an amplifier or a switch.  We will show in a later paper that free-carrier cavities can do much more than this, but that will build on the fundamentals discussed here.

The steady-state picture has been amply discussed in the literature \cite{Agrawal1979, Yurke2006, Kwon2013}, so it is not worth describing in more detail here.  Rather, we now proceed to look at the quantum noise and dynamics of these systems.

\subsection{Effective $\chi^{(3)}$ Model}

Next, we go from the steady-state picture to the limit of short carrier lifetime.  In this opposite limit, $\tau_{ph} \gg \tau_c$.  Typical devices do not realize this limit, but it is useful because it enables an apples-to-apples comparison between the free-carrier and Kerr effects.

To adiabatically eliminate the carrier number, one replaces $N_c$ with its steady-state value:

\beq
	N_c\,dt \rightarrow \frac{\eta(\alpha^*\alpha)dt + d\xi_N}{\gamma_{nr}}
\eeq
This gives the following SDE for the relevant dynamical variable, $\alpha$:

\beq
	d\alpha = \left[-\frac{\kappa + \eta}{2} - i\left(\Delta_c + \frac{\eta\delta_c}{\gamma_{nr}}(\alpha^*\alpha)\right)\right]\alpha\,dt - \sqrt{\kappa} d\beta_{in} + d\xi_\alpha' \label{eq:kerrsde}
\eeq
where the $d\xi_\alpha'$ is a new noise term that depends both on the $d\xi_\alpha$ and $d\xi_N$.  As before, the analogy to the Kerr model is clear: Equation~(\ref{eq:kerrsde}) is very close to the Wigner equations for the Kerr cavity \cite{Santori2014}, but the noise term is different.  The effect of this noise term will be discussed in the following sections, where the performance of Kerr- and free-carrier based amplifiers and switches is analyzed.

\section{Amplifier}
\label{sec:amp}

Figure \ref{fig:06-f2} shows that, for certain detunings, the state of the cavity changes very rapidly with a change in input power.  One can imagine using such a device to amplify differential signals: if the input signal is perturbed, that perturbation will be multiplied by some gain factor in the output.

\begin{figure}[tb]
\centering
\includegraphics[width=0.70\columnwidth]{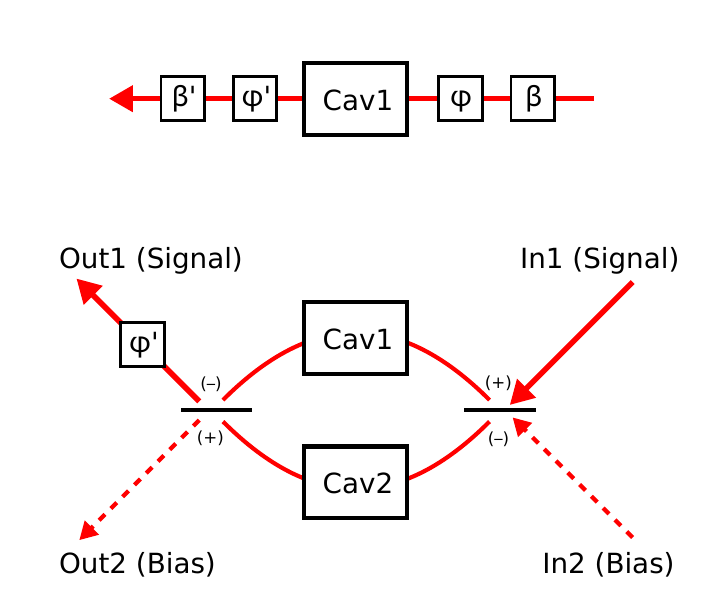}
\label{fig:06-f4}
\caption{(Color online) Left: Simple single-cavity amplifier, a cavity sandwiched between phase shifters and displacements $G = L(\beta') \triangleleft e^{i\phi'} \triangleleft \mbox{(Cav)} \triangleleft e^{i\phi} \triangleleft L(\beta)$.  Right: Symmetric two-cavity amplifier.}
\end{figure}

The real picture is actually a bit more complicated, since the input field has two quadratures.  In the Kerr cavity, one of the quadratures is amplified while the other is de-amplified \cite{Yurke2006}.  This gives rise to {\it phase-sensitive amplification} which, since there is no additional noise in the Kerr system, also squeezes the quantum noise of one quadrature below the vacuum level.

Key to an optical amplifier are its gain $G(\omega)$, its noise spectrum $S(\omega)$, and the scale on which nonlinear effects take over.  The gain and noise can be predicted by linearizing equations of motion (\ref{eq:sde-sc1}-\ref{eq:sde-sc2}) around the steady-state value.  This takes the general form:

\begin{eqnarray}
    d\bar{x} & = & \bar{A}\bar{x}\,dt + \bar{B}\,d\bar{\beta}_{\rm in} + \bar{F}\,dw \\
    d\bar{\beta}_{\rm out} & = & \bar{C}\bar{x}\,dt + \bar{D}\,d\bar{\beta}_{\rm in}
\end{eqnarray}
where $\bar{x}$ and $\bar{\beta}$ are {\it doubled-up} state vectors, which include the complex field operators and their conjugates \cite{Gough2010}, as well as the (real) carrier number: $\bar{x} = (\delta\alpha,\ \delta\alpha^*,\ \delta N_c)$, $d\bar{\beta} = (d\beta, d\beta^*)$ (removing any constant coherent input), and $\alpha$, $N_c$ are the steady-state values.

Linearization is key because many general results of stochastic systems theory only apply to linear or approximately linear systems \cite{AstromMurray}.  For example, in a linearized system, the output squeezing spectrum can be computed exactly for Gaussian inputs \cite{WallsMilburn, Crisafulli2013}.  Many results in quantum feedback control theory are also restricted to linear systems \cite{Nurdin2009, Hamerly2012}.

With a linearized model in hand, it is a simple matter to compute the internal state covariance $\bar{\sigma}$, the transfer and noise matrix $T(\omega)$, $N(\omega)$, and the frequency-domain input-output relation \cite{Gough2010, Hamerly2013}:

\begin{align}
	& \bar{A}\bar{\sigma} + \bar{\sigma}\bar{A}^\dagger + \frac{1}{2}\bar{B}\bar{B}^\dagger + \bar{F}\bar{F}^\dagger = 0 \label{eq:lyap} \\
	& \bar{\beta}_{{\rm out},\omega} = \underbrace{\left[\bar{D} + \bar{C} \frac{1}{-i\omega - \bar{A}}\bar{B}\right]}_{T(\omega)} \bar{\beta}_{{\rm in},\omega} +
		\underbrace{\bar{C} \frac{1}{-i\omega - \bar{A}} \bar{F}}_{N(\omega)} w_\omega \label{eq:dblt}
\end{align}

Unfortunately, because the doubled-up matrices here are 3-by-3 rather than 2-by-2, the analytic results are rather cumbersome and therefore not reproduced here.  Instead, in this section I compute these quantities numerically and compare the results to the Kerr system.  The results here are compared against a Kerr cavity with the same effective nonlinearity, $\chi = \eta\delta_c/\gamma_{nr}$.

\subsection{Gain}

\begin{figure}[tb]
\centering
\includegraphics[width=1.00\columnwidth]{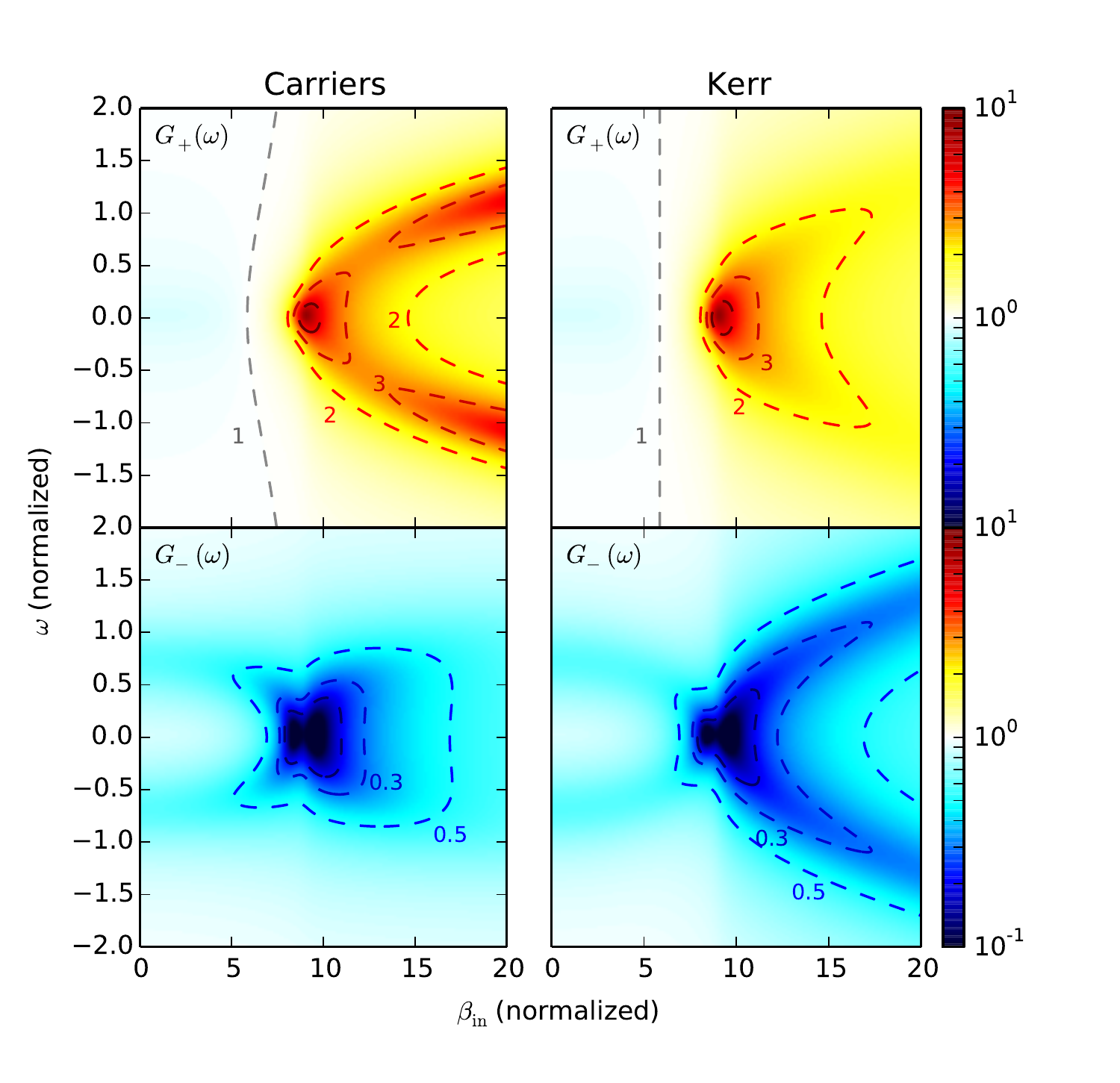}
\label{fig:06-f5}
\caption{(Color online) Plot of the maximum and minimum gain $G_+(\omega)$, $G_-(\omega)$ for free-carrier cavity (left) and Kerr cavity (right).  Parameters are from Table \ref{tab:t1}, with $\Delta_c = -0.7$.}
\end{figure}

The gain is computed from the singular values of the doubled-up transfer function $T(\omega)$.  If both singular values are the same, the device is a phase-insensitive amplifier.  Both the Kerr and free-carrier cavities, however, only amplify one quadrature.  As Figure~\ref{fig:06-f5} shows, they de-amplify the other quadrature as well.

At and below the ideal input $\beta_{\rm in} \approx 9$, the Kerr and free-carrier cavities seem to amplify in the same way.  For over-driven cavities, the behavior is very different.  The free-carrier cavity becomes very efficient at amplifying off-resonance, whereas the Kerr cavity hardly amplifies at all.

Gain is maximized when the system is very close to instability -- that is, when at least one of the eigenvalues of $\dbl{A}$ is very close to the imaginary axis.  From (\ref{eq:dblt}), an eigenvalue decomposition of $\dbl{A}$ gives the transfer function the following form:

\beq
	T = \dbl{D} + \sum_i \frac{v_i u_i^T}{-i\omega - \lambda_i}
\eeq
where $v_i, u_i$ are related to $B$, $C$ and the eigenvectors and $\lambda_i$ are the eigenvalues of $A$.  Since $B, C \sim O(\sqrt{\kappa}$), the numerator term is proportional to $\kappa$.  Near the resonance, the sum is dominated by the eigenvalue closest to zero, $\lambda_{\rm max}$.  The maximum gain should intuitively take the form:

\beq
	G \sim \frac{O(\kappa)}{|-i\omega - \lambda_{\rm max}|}
\eeq

This is a Lorentzian with a peak at $\mbox{Im}(\lambda_{\rm max})$ and bandwidth of $\Delta\omega = -\mbox{Re}(\lambda_{\rm max})$.  The peak gain is thus $G_{\rm max} = -O(\kappa)/(\mbox{Re}(\lambda_{\rm max}))$.  This gives us a gain-bandwidth relation:

\beq
	G_{\rm max} \Delta\omega = O(\kappa)
\eeq

The greater the amplifier gain, the slower it responds and the narrower its bandwidth.

\subsection{Internal State}

\begin{figure}[tb]
\begin{center}
\includegraphics[width=1.0\columnwidth]{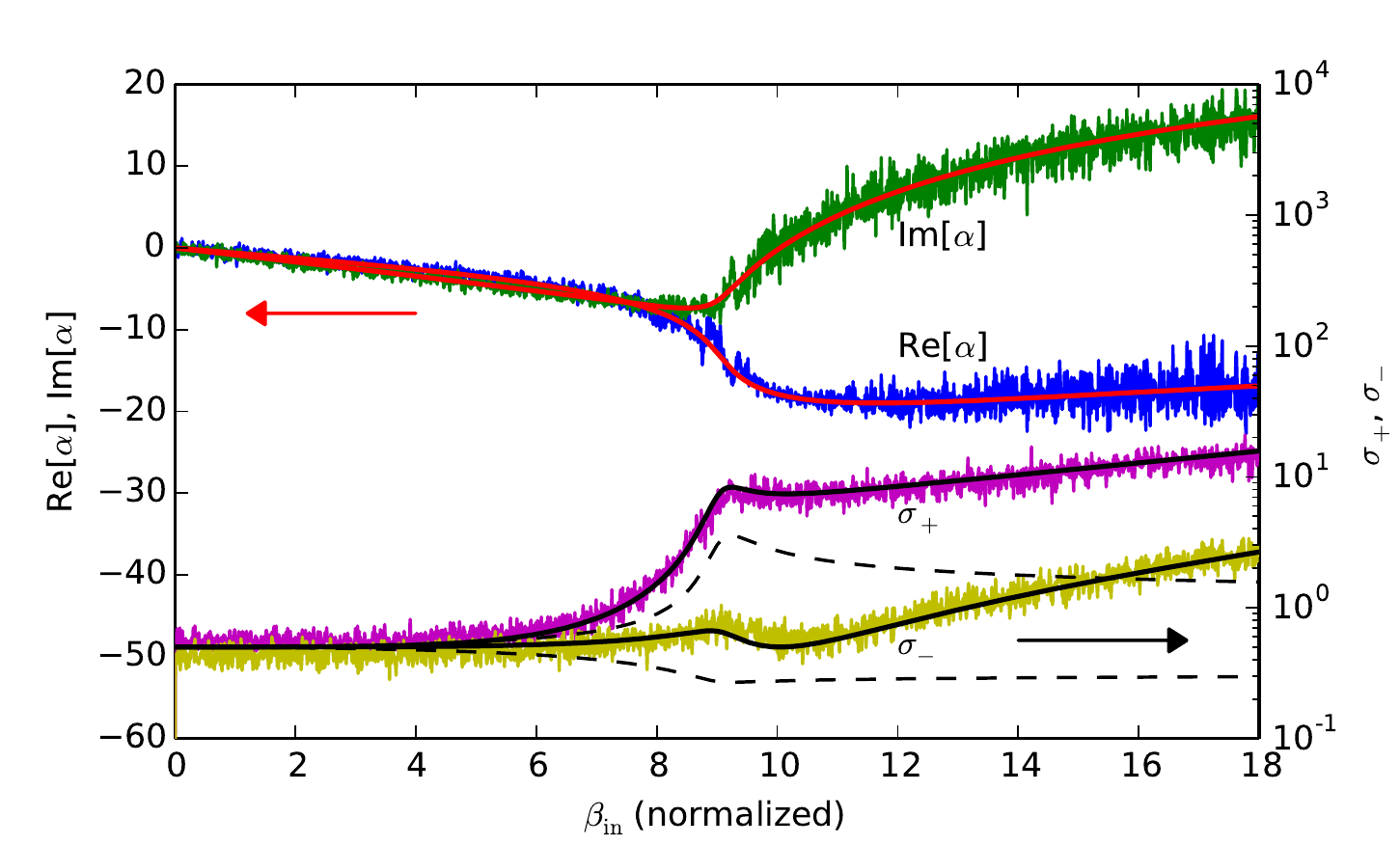}
\caption{(Color online) Internal state of free-carrier cavity, simulated eigenvalues of $\sigma$ ($\sigma_+$ and $\sigma_-$, the larger and smaller eigenvalue, respectively) compared to analytic result (solid lines).  The dashed line is the analytic result for an equivalent Kerr cavity.  $\Delta_c = -0.7$}
\label{fig:06-f7b}
\end{center}
\end{figure}

\begin{figure}[tb]
\begin{center}
\includegraphics[width=1.0\columnwidth]{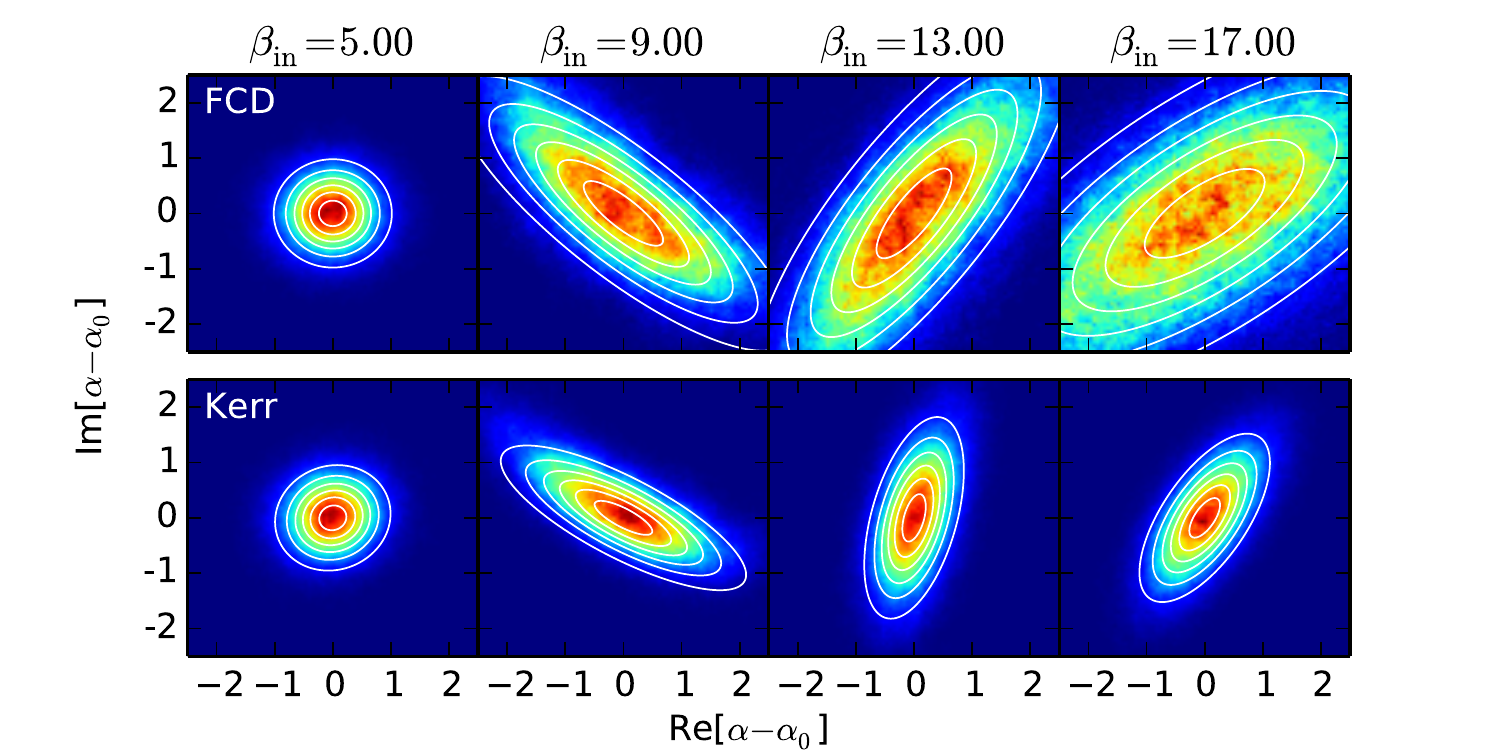}
\caption{(Color online) Simulated Wigner functions for free-carrier (top) and Kerr (bottom) cavities with the same effective $\chi^{(3)}$.  Analytic approximation for linearized model given in white contours.  $\Delta_c = -0.7$}
\label{fig:06-f7a}
\end{center}
\end{figure}

The internal state is computed using the Lyapunov equation (\ref{eq:lyap}).  This time, the Wigner equations contain additional noise terms, which make the state noisier than the state of an equivalent Kerr cavity.  This is plotted in the Figures~\ref{fig:06-f7b}-\ref{fig:06-f7a}.  The state remains roughly Gaussian, but the size of the Gaussian is larger than in the Kerr case, especially above the inflection point.

Unlike in the Kerr case, the mode in the free-carrier cavity is never squeezed.  As seen in Figure~\ref{fig:06-f7b}, the eigenvalues $\sigma_+, \sigma_-$ of the covariance matrix $\sigma$ are always $\geq \frac{1}{2}$, ensuring that the state is always ``classical'' in the sense that it has a valid $P$ representation.  Given that the carrier excitation and decay process is highly incoherent, it should not be too surprising that the cavity always remains in a classical state.  But it is a clear departure from the Kerr model, and this classicality could conceivably be used to distinguish between the two in an experiment.

Also note that the noise grows linearly with the input field at high powers.  This happens because the free-carrier number is constantly fluctuating, being driven by excitation and decay events that mimic a Poisson process.  At high carrier numbers, this means that the cavity detuning and consequently the cavity field become very noisy.  This does not happen in the Kerr cavity, where the nonlinearity is mediated by virtual transitions which do not add any noise to the system.  It is a peculiar consequence of the incoherence of the free-carrier mechanism.

\subsection{Output Noise Spectrum}

\begin{figure}[tbp]
\begin{center}
\includegraphics[width=1.00\columnwidth]{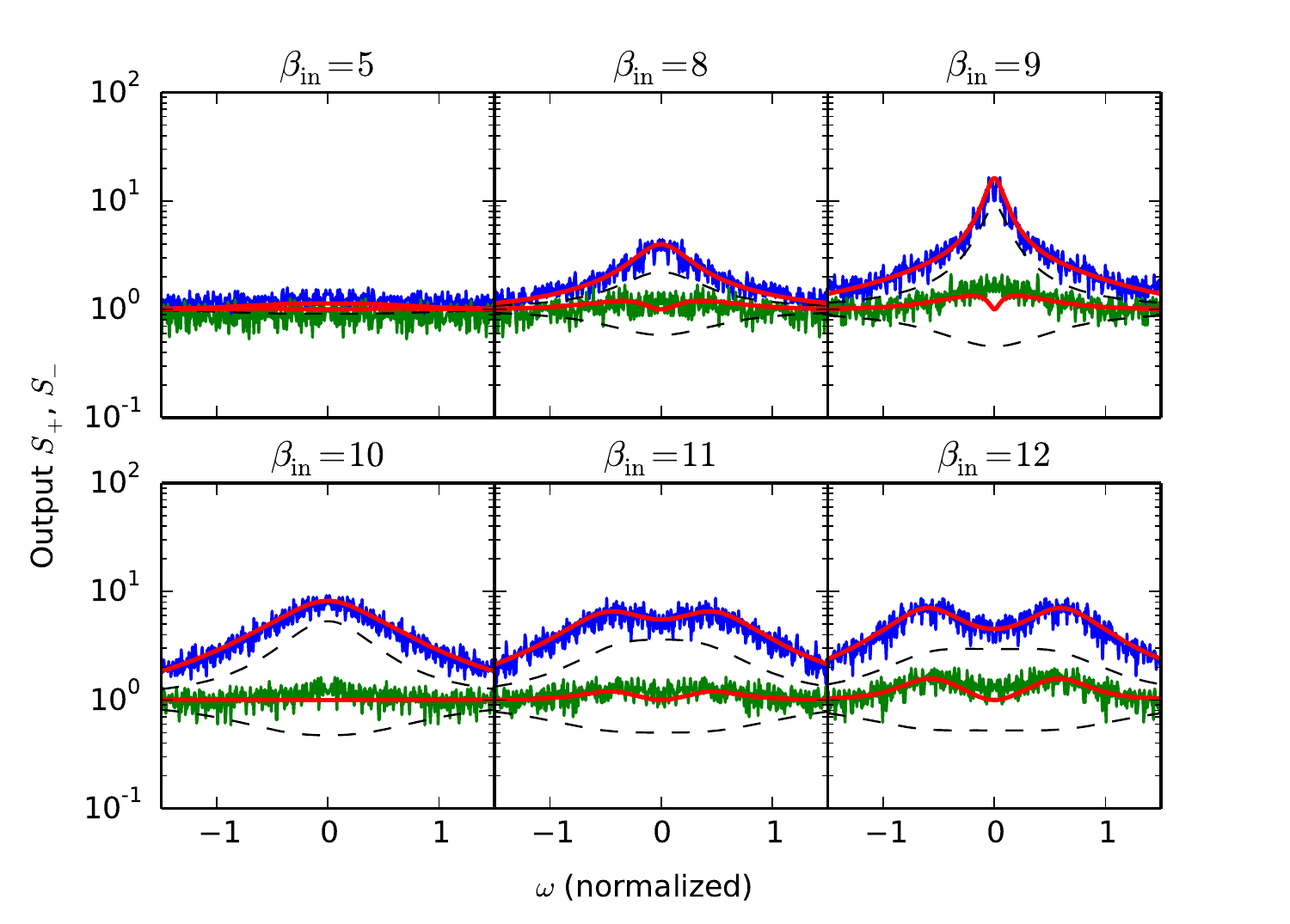}
\caption{(Color online) Noise spectrum modes $S_+$, $S_-$ for the a free-carrier cavity with $\Delta_c = -0.7$ at various pump powers.  Green and blue lines are numerical simulations; red solid line is the prediction from the linearized ABCD model.  The dashed lines are the prediction from the Kerr model.}
\label{fig:06-f8}
\end{center}
\end{figure}

\begin{figure}[tbp]
\begin{center}
\includegraphics[width=1.0\columnwidth]{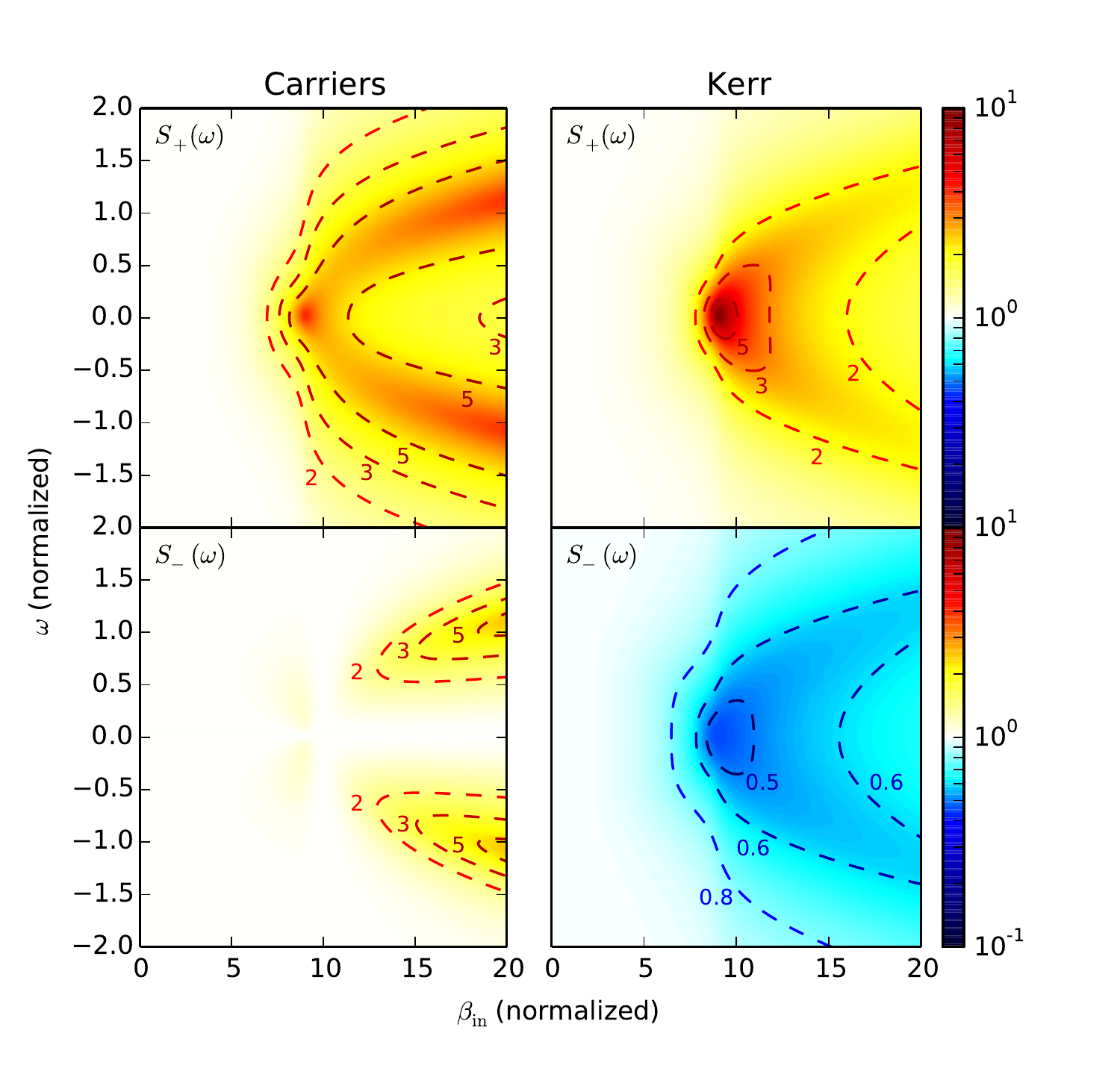}
\caption{(Color online) Noise spectrum $S_+$, $S_-$ as a function of input $\beta_{\rm in}$ and frequency $\omega$ for Kerr and free-carrier models.  $\Delta_c = -0.7$}
\label{fig:06-f9}
\end{center}
\end{figure}

Given a linearized input-output model, we can compute the squeezing spectrum (noise spectrum) for the cavity output field \cite{WallsMilburn, Gough2009Sq}.  The squeezing spectrum for quadrature $\theta$ is defined as the power spectral density of a homodyne measurement of $\beta_{\rm out}(t)$.  That is, for the following homodyne signal,

\beq
	j_\theta(t) = e^{-i\theta}\beta_{\rm out}(t) + e^{i\theta} \beta_{\rm out}^*(t)
\eeq
the squeezing spectrum is:

\beq
	S_\theta(\omega) = \sqrt{2 P_\theta(\omega)},\ \ P_\theta(\omega) = \frac{\avg{j_\theta(\omega)^*j_\theta(\omega')}}{2\pi\delta(\omega-\omega')}
\eeq

$S_{\theta}(\omega)$ is normalized so that the coherent state has $S_\theta(\omega) = 1$.  For general states, $S_\theta(\omega)$ depends on $\theta$.  The maximum and minimum of $S_\theta(\omega)$, with respect to $\theta$, are denoted $S_+(\omega)$ and $S_-(\omega)$, respectively.

The squeezing spectrum of the Kerr cavity can be computed analytically \cite{Yurke2006}.  By contrast, since the free-carrier squeezing spectrum involves the inverse of a $3\times3$ matrix, it is unlikely that a simple expression can be found.  However, it is not difficult to compute numerically.

In Figure~\ref{fig:06-f8}, the noise spectrum is obtained in two separate ways: first, simulating the full system in the time domain and taking the Fourier transform of the homodyned output (blue, green curves); and second, from the analytic predictions of the linearized ABCD model.  These agree everywhere except for very large pump powers, where the system approaches a bifurcation.

Figure~\ref{fig:06-f9} displays the noise spectrum for the whole range $0 \leq \beta_{\rm in} \leq 20$, for both Kerr and free-carrier devices.  Two things are obvious.  First, the noise curve (at least for the $S_+$ component) matches the general form of the gain curve in Figure~\ref{fig:06-f5}.  This is of course necessary because there must be noise wherever there is gain.  The free-carrier cavity, unlike the Kerr cavity, amplifies not only at $\omega = 0$ near the point of maximum gain, but also for $\omega \neq 0$ for $\beta_{\rm in}$ above that point.

Unlike the Kerr cavity, the free-carrier cavity does not squeeze the output field.  Regardless of the parameters, regardless of the pump power, both $S_+$ and $S_-$ are always above the vacuum level, indicating that this is a classical field with no squeezing.  The Kerr cavity, on the other hand, squeezes light over a broad range of the spectrum.  This is in agreement with the results of the previous section, which showed that the internal field of the free-carrier cavity was classical.  If the input and intracavity field are in a classical state, so is the output.

\section{Latch}
\label{sec:latch}

\begin{figure}[tbp]
\begin{center}
\includegraphics[width=0.80\columnwidth]{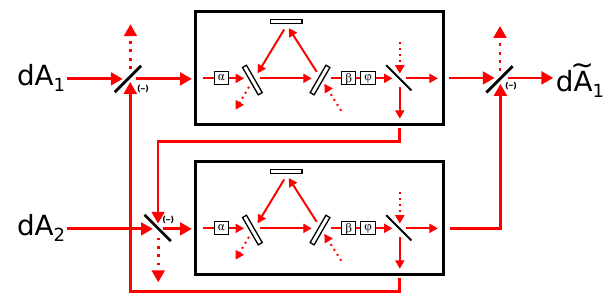}
\caption{(Color online) Circuit diagram for a photonic SR-Latch}
\label{fig:06-f15}
\end{center}
\end{figure}

It is also possible to construct a switching device using only amplifiers, provided the amplification is large enough \cite{Mabuchi2011}.  The circuit in Figure~\ref{fig:06-f15} uses two identical amplifiers in a feedback loop.  Suppose that each amplifier has a gain $G$.  Consider the fate of a perturbation in the top amplifier.  An input $\delta\beta$ is amplified to $G\,\delta\beta$.  This amplifier has a fan-out of 2, so $(G/\sqrt{2})\delta\beta$ passes to the right and exits the system, while $(G/\sqrt{2})\delta\beta$ passes to the lower amplifier.

In the lower amplifier, it grows to $(G^2/2)\delta\beta$, is fed back into the original amplifier.  After passing through this loop, the signal strength has grown to $(G^2/2)\delta\beta$.  This leads to a latching instability if the gain is sufficiently large:

\beq
	G > \sqrt{2}
\eeq

Symmetry gives the latch some very desirable properties.  Unlike the single-cavity switch, the two states here are symmetric.  Thus, there is less worry about finding the right bias field to ``balance'' the low and high state, and transitions between the states look the same.  But this comes at the cost of the added complexity of two cavities, plus the extra connections.

\begin{figure}[tbp]
\begin{center}
\includegraphics[width=1.0\columnwidth]{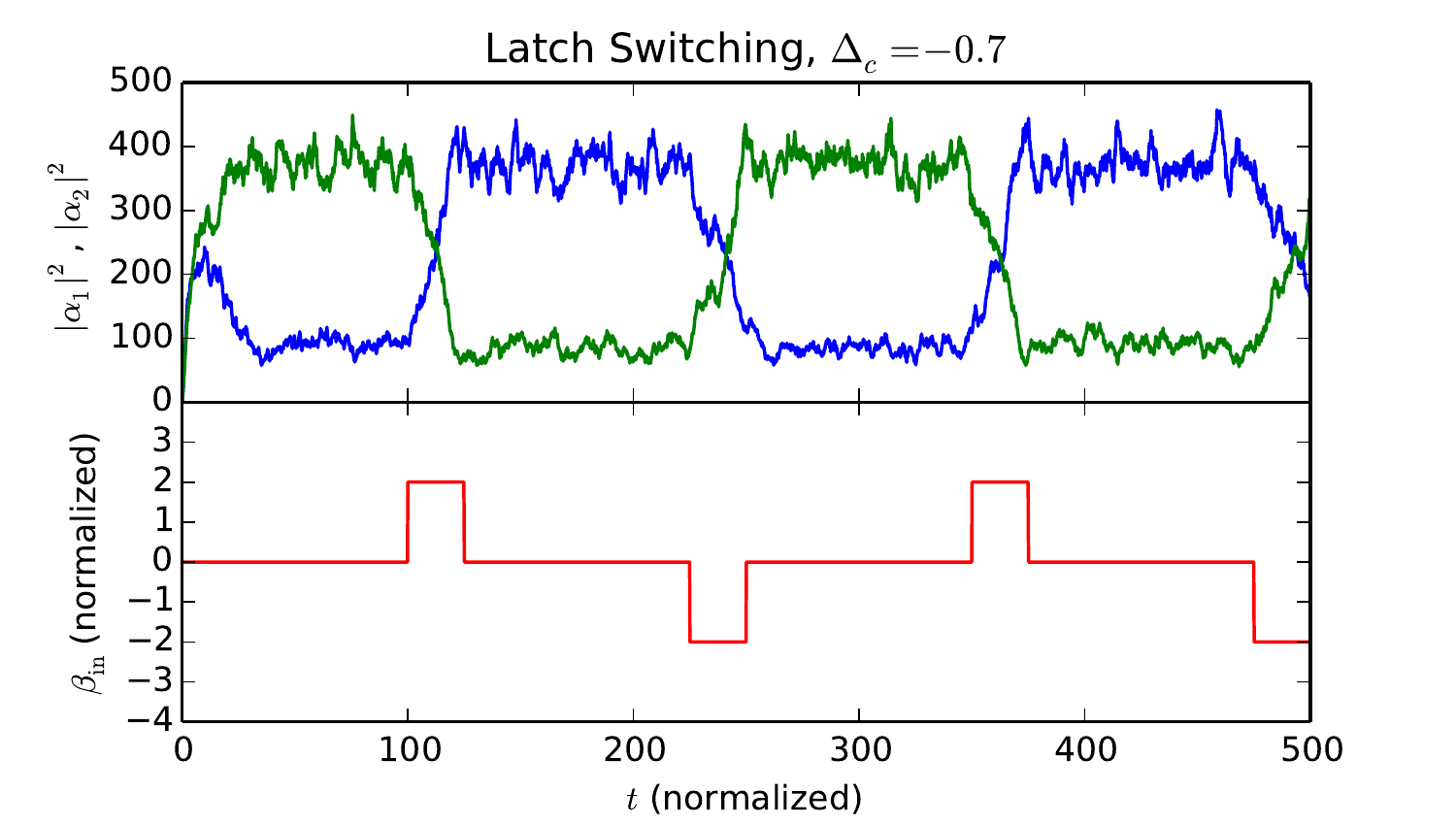}
\caption{(Color online) Top: time series of the latch internal state.  Bottom: input field.}
\label{fig:06-f16}
\end{center}
\end{figure}

Figure \ref{fig:06-f16} shows a latch simulation for the same parameters used in the previous section.  Here, the detuning is set to $\Delta_c = -0.7$, large enough to realize a large gain, but not large enough make an individual cavity bistable.  The symmetry between the two states is very clear.

\begin{figure}[tbp]
\begin{center}
\includegraphics[width=1.0\columnwidth]{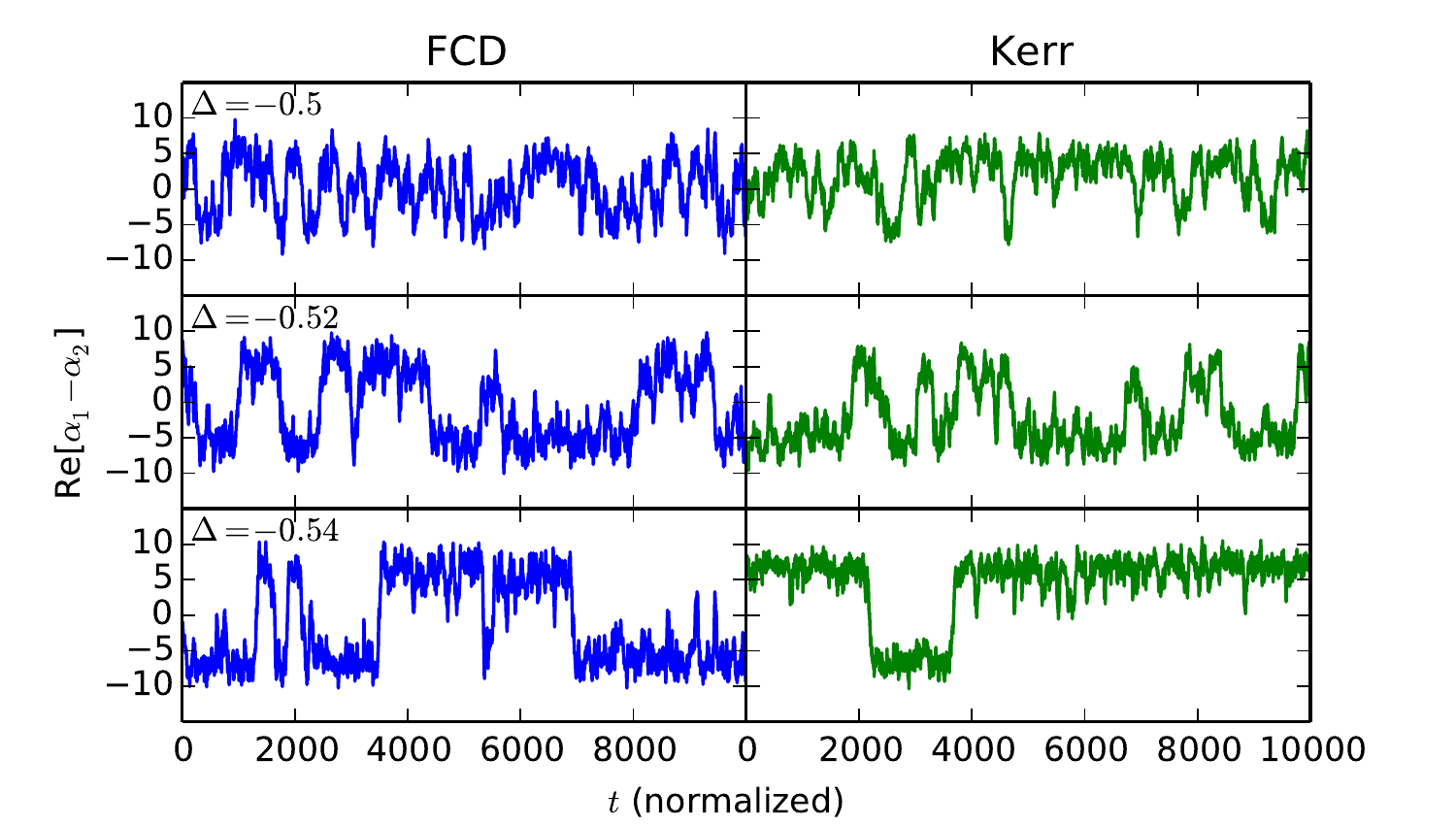}
\caption{(Color online) Asymmetric part of the latch state Re[$\alpha_1 - \alpha_2$] for free-carrier based latch (left) and Kerr-based latch of the same $\chi^{(3)}$.  Cavity detuning set to $\Delta_c = 0.50, 0.52, 0.54$.}
\label{fig:06-f17}
\end{center}
\end{figure}

\begin{figure}[tbp]
\begin{center}
\includegraphics[width=1.0\columnwidth]{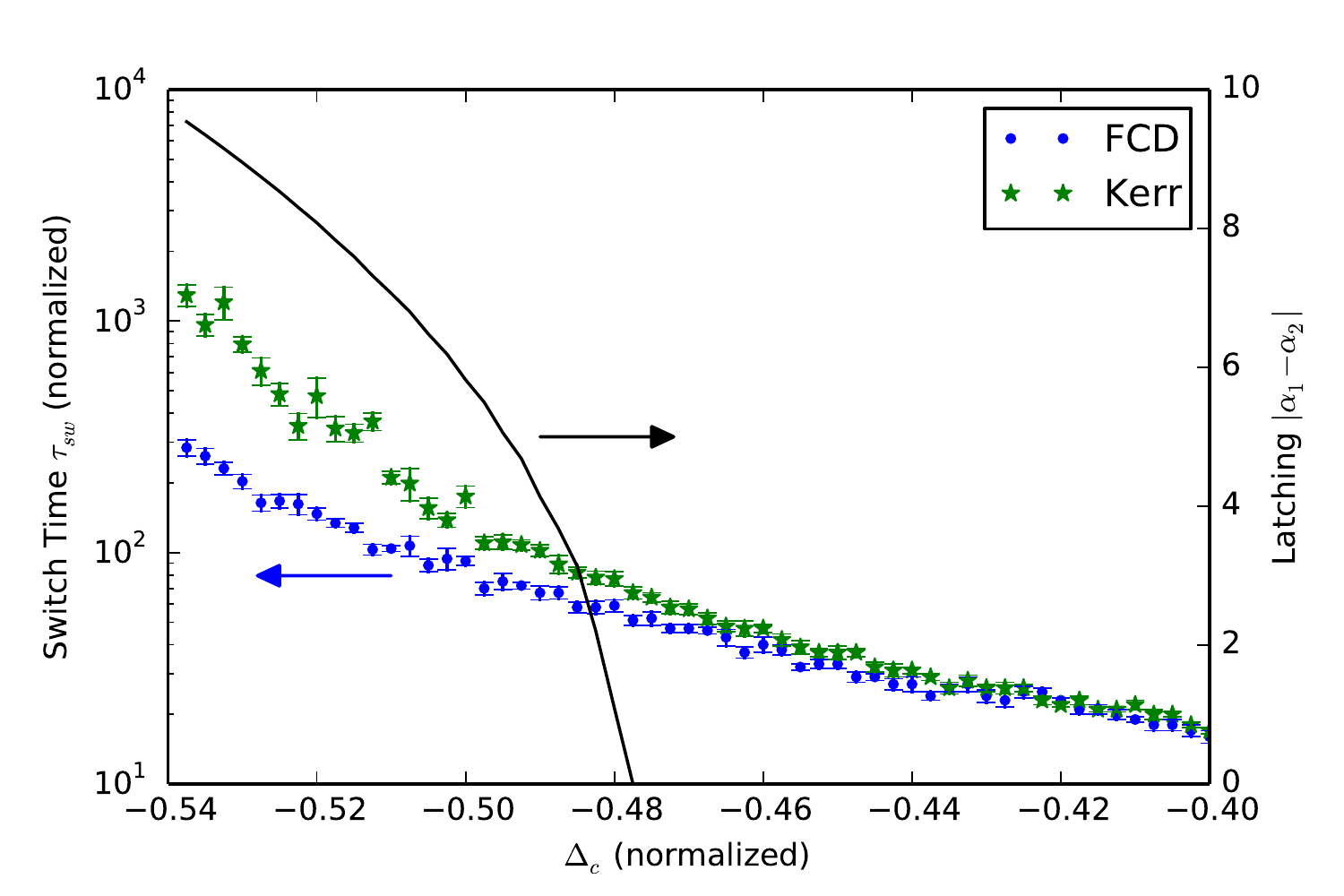}
\caption{(Color online) Spontaneous switching lifetimes $\tau_{sw}$ for the Kerr and free-carrier latch as a function of detuning.}
\label{fig:06-f18}
\end{center}
\end{figure}

Externally driven switching in the latch is {\it good}, because it allows the user to set the state of the latch, which becomes a memory element.  But thanks to quantum noise, Kerr and free-carrier devices also undergo {\it spontaneous switching}.  This is generally {\it bad}, because it limits the lifetime of a carrier-based memory.

In the Kerr case, spontaneous switching is driven by vacuum fluctuations \cite{Santori2014}.  In the free-carrier case, vacuum fluctuations combine with stochastic carrier excitation and decay to drive the switching process.  Because there are more fluctuations, we naturally expect the free-carrier cavity to spontaneously switch at a higher rate than the Kerr cavity.

When the switching rate is low, the switching process is well described by a two-state Markov chain.  In a two-state Markov chain, there are two states $a$ and $b$, with jump probabilities

\beq
	P(a \rightarrow b) = \gamma_{a} dt,\ \ \
	P(b \rightarrow a) = \gamma_{b} dt
\eeq

In the latch, the states are symmetric, so $\gamma_a = \gamma_b \equiv \gamma_{sw}$.  The probability of being in a given state evolves as:

\bea
	\frac{dP_a}{dt} & = & -\gamma_{sw} P_a + \gamma_{sw} P_b \nonumber \\
	\frac{dP_b}{dt} & = & \gamma_{sw} P_a - \gamma_{sw} P_b
\eea

Solving this linear system, one finds that the system reverts to its equilibrium distribution with a characteristic time $\tau_{sw} = 1/(2\gamma_{sw})$.  This time can be measured from simulations of the latch by looking at the autocorrelation function $R(\tau)$, which decays exponentially for the Markov process:

\beq
	R(\tau) = \frac{\avg{\alpha(t)\alpha(t-\tau)^*}}{\avg{\alpha(t)\alpha(t)^*}} \rightarrow e^{-\tau/\tau_{sw}}
\eeq

Figure \ref{fig:06-f17} shows time traces of the asymmetric field $\alpha_1 - \alpha_2$ as the latch detuning is varied from $-0.50$ to $-0.54$, about where the latching transition occurs.  Larger negative detunings correspond to higher gain (see Fig.~\ref{fig:06-f2}), and likewise stronger latching.  However, for a fixed detuning, the free-carrier cavity has a shorter spontaneous switching lifetime.

This is also seen in Figure \ref{fig:06-f18}, which plots $\tau_{sw}$ for the free-carrier and equivalent Kerr latches.  Because the free-carrier cavity has more quantum noise than the Kerr cavity, its spontaneous switching rate is higher.  The effect becomes noticeable once the latching transition sets in, and grows as the latching grows stronger.

\section{Conclusion}
\label{sec:conclusion}

In this paper, we introduced a method to simulate optical cavities where free-carrier dispersion is the dominant nonlinearity.  This method is based on deriving an approximate Fokker-Planck equation for the Wigner function, the approximation being valid in the weak-coupling limit where the detuning per carrier is much smaller than the cavity linewidth and the mean photon number is large.  Importantly, this allows us to keep track of the dominant quantum effects (vacuum noise in the optical field, Poisson noise in the carrier excitation and decay) without running a full quantum simulation.  This method was then applied to simulate an optical amplifier and an SR-latch.

Because the semiclassical properties of these devices are well known, in this paper we focused on the quantum noise in the free-carrier amplifier and latch.  Since the free-carrier dispersion creates an effective $\chi^{(3)}$ nonlinearity, in both cases the free-carrier simulations were compared to simulations of an analogous Kerr-based device, for which the quantum model is well known.  In the semiclassical, steady-state limit, the Kerr and free-carrier devices behave the same.

For the phase-sensitive amplifier, we find that the free-carrier and Kerr devices have the same gain near the amplification maximum.  However, the noise in the amplification direction is larger for most frequencies.  In the other quadrature, the Kerr cavity squeezes the field, while the free-carrier cavity does not produce any squeezing.  In the SR-latch, we notice a difference in the the spontaneous switching rate: the free-carrier latch has switching rates of $\gtrsim 5$ larger than the Kerr latch, a discrepancy that grows as the latching becomes stronger.

\begin{acknowledgements}

The authors would like to thank Nikolas Tezak and Charles Santori for helpful discussions.  This work has been supported by DARPA-MTO under award N66001-11-1-4106.  RH is supported by a Stanford Graduate Fellowship.

\end{acknowledgements}

\appendix

\section{Closedness of Operator Algebra}
\label{sec:closedness}

\subsection{Closedness and the Wigner Function}

The single-mode model of Sec.~\ref{sec:singlemode} reduces the number of phase-space dimensions from $5N+2$ to $7$, but the operator algebra $X = [a, a^\dagger, \sigma_+, \sigma_-, n, \bar{n}, Q]$ is a very restricted basis set.  Many degrees of freedom cannot be expressed in terms of the $X_i$.  However, if certain closedness conditions are satisfied, operators in the algebra stay in the algebra under time evolution.  Since the Wigner function is tied to expectations of operator products, this allows us to set up a PDE for the Wigner function.  In essence, the degrees of freedom contained in the single-mode Wigner function exactly ``decouple'' from the other degrees of freedom in the system, and the single-mode model is valid.

$X$ is closed under commutation and thus forms a valid basis for an algebra $\mathcal{B}$ -- a vector space spanned by the $X_i$ and their products, e.g.\ $X_i X_j, X_i X_j X_k$, etc.  We say that $\mathcal{B}$ is {\it closed under time evolution} if the time derivative of every element of $\mathcal{B}$ is in $\mathcal{B}$:

\beq
	B \in \mathcal{B}\ \ \Rightarrow\ \ \left.\frac{dB}{dt}\right|_{\rm ad} \in \mathcal{B} \label{eq:closed}
\eeq

If (\ref{eq:closed}) holds, then for every c-number product $x_m\ldots x_p$, there exists a polynomial $M_{i\ldots p}(x)$ such that

\beq
	\left(\frac{d(X_m \ldots X_p)_{sym}}{dt}\right)_{\rm p} = M_{m\ldots p}(x)
\eeq
and from the correspondence between Wigner moments and operator products, we obtain an equation of motion for the Wigner function's moments:

\beq
	\frac{d}{dt}\langle x_m \ldots x_p \rangle_{W} = \langle M_{m\ldots p}(x) \rangle_{W} \label{eq:moments}
\eeq

It is a well-known result in stochastic calculus that we can recast (\ref{eq:moments}) as a generalized Fokker-Planck equation for $W(x, t)$, where the moments are replaced by cumulants.  Eqs.~(\ref{eq:fpe-sde}-\ref{eq:cum2}) arise when this equation is truncated to second order.

\subsection{Proof of Closedness}

We will prove closedness for a relatively broad class of Hamiltonians and Lindblad terms.  To start, define a boson space $\mathcal{B}_\k$, a restricted fermion space $\mathcal{F}_\k$, and sum-operator spaces $\mathcal{B}^{(n)}$ (note no index $\k$):

\bea
	\mathcal{B}_\k & \equiv & \mbox{span}(\sigma_{+\k}, \sigma_{-\k}, n_{\k}, \bar{n}_{\k}, Q_{\k}) \\
	\mathcal{F}_\k & \equiv & \{x_1 c_\k + x_2 \bar{c}_\k^\dagger\ |\ x_1,x_2\in \mathcal{B}_\k\} \\
	\mathcal{B}^{(n)} & \equiv & \mbox{span}(X_{i_1}\ldots X_{i_m}, m \leq n),\ \ \mathcal{B} \equiv \mathcal{B}^{\infty}
\eea
For example, $c_\k \in \mathcal{F}_\k$, $Q \in \mathcal{B}^{(1)}$, $\sigma_+^2 \in \mathcal{B}^{(2)}$.  Below, we prove several lemmas about the ordering of bosonic and fermionic operators.

\theorem{Lemma 1.}  If $f \in \mathcal{F}_\k$ and $b \in \mathcal{B}^{(1)}$, then $f b = b f + f'$ and $f^\dagger b = b f^\dagger + (f'')^\dagger$, where $f', f'' \in \mathcal{F}_\k$.

\proof The case for $f$ is proved by a search of all relevant cases.  $\mathcal{B}$ and $\mathcal{F}_\k$ have 7 and 10 basis vectors, respectively, so this is 70 commutators to check (most are zero).  Given this, the $f^\dagger$ case holds because $[f^\dagger, b]^\dagger = -[f, b^\dagger]$.

\theorem{Lemma 2.}  If $f \in \mathcal{F}_\k$ and $b \in \mathcal{B}^{(n)}$, then $f b = b f + \sum_i b'_i f'_i$ and $f^\dagger b = b f^\dagger + \sum_i b''_i (f''_i)^\dagger$, where $f_i', f_i'' \in \mathcal{F}_\k$ and $b_i', b_i'' \in \mathcal{B}^{(n-1)}$.

\proof Induction on $n$.  The $n=1$ case is proved in Lemma 1.  Assuming it holds for $n-1$, write $b = \sum_j b_{1,j} b_{n-1,j}$ with $b_{1,j} \in \mathcal{B}^{(1)}, b_{n-1,j} \in \mathcal{B}^{(n-1)}$.  Using both Lemma 1 and the $n-1$ case, we move the fermionic operator from the left to the right side of the expression (summation signs omitted for claity):

\bea
	f b & = & f b_{1,j} b_{n-1,j} = (b_{1,j} f + f'_j) b_{n-1,j} \nonumber \\
	& = & b_{1,j}(b_{n-1,j} f + b'_{n-2,jk} f''_{jk}) + (b_{n-1,j} f'''_j + b''_{n-2,jk} f''''_{jk}) \nonumber \\
\eea
With appropriate index renaming, this takes the desired form.  The $f^\dagger$ case is analogous.
	
\theorem{Lemma 3.}  If $f_\k, f'_\k \in \mathcal{F}_\k$, then $f_\k^\dagger f'_\k \in \mathcal{B}_\k$ and $\sum_\k f_\k^\dagger f'_\k \in \mathcal{B}^{(1)}$.

\proof Done by a search of all relevant cases -- 100 in all since $\mathcal{F}_\k$ has 10 basis vectors.  

\theorem{Theorem 1.}  The operator algebra $\mathcal{B}$ is closed under (\ref{eq:05-ad}) if the Hamiltonian is in $\mathcal{B}$ and the Lindblad terms take the following form: $L \sim b f_{\k_1} \ldots f_{\k_n}$, with $b \in \mathcal{B}$ and either $f_{\k_i} \in \mathcal{F}_{\k_i}$ or $f_{\k_i}^\dagger \in \mathcal{F}_{\k_i}$, or $L \sim b g_\k$, for $b \in \mathcal{B}, g \in \mathcal{B}_\k$.  There must be one $L$ for each multi-index $\k_i$.

\proof To prove closedness, we must show that (\ref{eq:05-ad}) is in $\mathcal{B}$ for all $A\in\mathcal{B}$.  This is the sum of a Hamiltonian and Lindblad terms.  The Hamiltonian term $-i[A, H]$ is obvious since both $A$ and $H$ are in the algebra of $X$, which is closed under commutation.

The Lindblad term is $\frac{1}{2}(2L^\dagger A L - L^\dagger L A - A L^\dagger L)$.  We first use Lemma 2 to move the indexed parts $f_{\k_1} \ldots f_{\k_n}$ to the same side of the expression; for instance, for $L^\dagger A L$, we find

\begin{align}
	& (f'_{\k_n})^\dagger \ldots (f'_{\k_1})^\dagger b^\dagger A b f_{\k_1} \ldots f_{\k_n} \nonumber \\
	& \rightarrow \sum_{i} b'_{i_1\ldots i_n, j_1\ldots j_n} (f'_{i_n,\k_n})^\dagger \ldots (f'_{i_1,\k_1})^\dagger f_{j_1,\k_1} \ldots f_{j_n,\k_n}
\end{align}
To each term in this sum, we apply Lemma 3 to combine the fermionic operators into bosonic operators.  

\begin{align}
	& \sum_{\k_1} (f'_{i_n,\k_n})^\dagger \ldots (f'_{i_1,\k_1})^\dagger f_{j_1,\k_1} \ldots f_{j_n,\k_n} \nonumber \\
	& \ \ \ \rightarrow (f'_{i_n,\k_n})^\dagger \ldots (f'_{i_2,\k_2})^\dagger b_1 f_{j_2,\k_2} \ldots f_{j_n,\k_n}
\end{align}
Summation over $\k$ is critical here; without it $b_1$ would not be a bosonic sum-operator in $\mathcal{B}^{(1)}$.  Thus, the algebra $\mathcal{B}$ is not closed for a Lindblad term with just a single $\k$ -- we must sum over all the $\k$'s in order to recover closedness.

Now we use Lemma 3 to move the bosonic operator $b$ to the left, recombine the operators with index $\k_2$, and repeat until all fermionic operators have been combined.  This gets rid of all the indices $\k_i$, resulting in an operator that lives in $\mathcal{B}$.  The terms $L^\dagger L A$ and $A L^\dagger L$ are done the same way.  It follows that the Lindblad term in (\ref{eq:05-ad}) lives in $\mathcal{B}$.  As before, the action of a single Lindblad term breaks closedness, but when we sum over $\k$, it is recovered.

The result for $L = b g_\k$ can be shown without Lemmas 1--3.  We just use the commutation relations of the $X_\k$ to move the all the indexed terms to the same side, where they can be combined and summed into a term in $\mathcal{B}$.

This theorem encompasses all the quantum models studied in this paper.  A few examples of things it does {\it not} apply to would be index-dependent effects, say $H \sim E_\k n_\k$, or certain effects that violate charge conservation, such as $L \sim c_\k \bar{c}_\k^\dagger$.

\newpage

\section{Full Wigner SDE's}
\label{sec:fullsdes}

The cumulants $C^{(1)}$ and $C^{(2)}$ are computed from Eqs. (\ref{eq:cum1}-\ref{eq:cum2}) using {\it Mathematica}.  The terms are separated by physical origin in the sections below.

\subsection{Uncoupled Cavity, Carrier Terms}

In this case, $H = \Delta_c a^\dagger a + \Delta_e(n + \bar{n})/2$ and $L + \sqrt{\kappa}\,a$.  It is easy to show that:

\beq
	C^{(1)} = \begin{bmatrix} (-i\Delta_c-\frac{\kappa}{2})\alpha \\ (i\Delta_c-\frac{\kappa}{2})\alpha^* \\
	-i\Delta_e v \\ i\Delta_e \bar{v} \\ 0 \\ 0 \\ 0 \end{bmatrix},\ \ \
	C^{(2)} = \begin{bmatrix}
		 0 & \frac{\kappa}{2} & 0 & 0 & 0 & 0 & 0 \\
		\frac{\kappa}{2} & 0 & 0 & 0 & 0 & 0 & 0 \\
		 0 & 0 & 0 & 0 & 0 & 0 & 0 \\
		 0 & 0 & 0 & 0 & 0 & 0 & 0 \\
		 0 & 0 & 0 & 0 & 0 & 0 & 0 \\
		 0 & 0 & 0 & 0 & 0 & 0 & 0 \\
		 0 & 0 & 0 & 0 & 0 & 0 & 0
	\end{bmatrix} \label{eq:app1}
\eeq

\begin{widetext}

\subsection{Photon-Carrier Interaction}

Here, $H = ig (a^\dagger \sigma_{-} - a\sigma_{+})$, and there are no environment couplings.  There is no noise term here.

\beq
	C^{(1)} = \begin{bmatrix} gv \\ gv^* \\ -g\alpha(N-m-\bar{m}) \\ -g\alpha^*(N-m-\bar{m}) \\
		-g(\alpha v^* + v\alpha^*) \\ -g(\alpha v^* + v\alpha^*) \\ -g(\alpha v^* + v\alpha^*)
		\end{bmatrix},\ \ \
	C^{(2)} = 0
\eeq

\subsection{Free-Carrier Dispersion / Absorption}

Here, $L_\k = \sqrt{\gamma_{fca}}\,n_\k a, \sqrt{\bar{\gamma}_{fca}}\,n_\k a$ and $H = a^\dagger a(\delta_{fcd} n + \bar{\delta}_{fcd} \bar{n})$.  Define $\delta_{fc} = \delta_{fcd} - i\gamma_{fca}/2$.  Then the cumulants become:

\beq
	C^{(1)} = \begin{bmatrix}
		-i(\delta_{fc}m+\bar{\delta}_{fc}\bar{m})\alpha \\
		i(\delta_{fc}^*m+\bar{\delta}_{fc}^*\bar{m})\alpha^* \\
		-i (\delta_{fc}+\bar{\delta}_{fc}) (\alpha^*\alpha - \frac{1}{2}) v \\
		i (\delta_{fc}^*+\bar{\delta}^*_{fc}) (\alpha^*\alpha - \frac{1}{2}) v^* \\ 0 \\ 0 \\ 0 \end{bmatrix},\ \ \
	C^{(2)} = \begin{bmatrix}
		0 & (\gamma_{fca}m+\bar{\gamma}_{fca}\bar{m})/2\!\!\!\!\!\!\!\! & 0 & \!\!\!\!\!\!\!\!(\gamma_{fca}+\bar{\gamma}_{fca})v^*\alpha\!\!\!\!\!\!\!\! & 0 & 0 & 0 \\
		* & 0 & \!\!\!\!\!\!\!\!(\gamma_{fca}+\bar{\gamma}_{fca})v\alpha^*\!\!\!\!\!\!\!\! & 0 & 0 & 0 & 0 \\
		0 & * & 0 & \!\!\!\!\!\!\!\!\!\!\!\!\!\!\begin{array}{c}(\gamma_{fca}+\bar{\gamma}_{fca})(\alpha^*\alpha-\frac{1}{2})\\\times\ (N+2q-m-\bar{m})\end{array}\!\!\!\!\! & 0 & 0 & 0 \\
		* & 0 & * & 0 & 0 & 0 & 0 \\
		0 & 0 & 0 & 0 & 0 & 0 & 0 \\
		0 & 0 & 0 & 0 & 0 & 0 & 0 \\
		0 & 0 & 0 & 0 & 0 & 0 & 0
		\end{bmatrix} \label{eq:fca-c12}
\eeq

Equation~(\ref{eq:fca-c12}) has a nontrivial noise matrix.  However, this can be greatly simplified in the non-degenerate, fast-dephasing limit usually taken.

\subsection{Recombination}

This is mediated by the term $L = \sqrt{\gamma_{rc}}\,\sigma_-$.  Recombination only takes place when an electron and hole occupy the same state $\k$, so the rate goes as the pair density $q$, not as the carrier density $m+\bar{m}$.

\beq
	C^{(1)} = \gamma_{rc} \begin{bmatrix} 0 \\ 0 \\ -\frac{1}{2}v \\ -\frac{1}{2}v^* \\ -q \\ -q \\ -q
		\end{bmatrix},\ \ \
	C^{(2)} = \begin{bmatrix}
		0 & 0 & 0 & 0 & 0 & 0 & 0 \\
		0 & 0 & 0 & 0 & 0 & 0 & 0 \\
		0 & 0 & 0 & \!\!\!\!\!\!\!\!\!\!\frac{1}{2}(N+2q-m-\bar{m}) & \frac{1}{2} v & \frac{1}{2} v & \frac{1}{2} v \\
		0 & 0 & \frac{1}{2}(N+2q-m-\bar{m})\!\!\!\!\!\!\!\!\!\! & 0 & \frac{1}{2}v^* & \frac{1}{2}v^* & \frac{1}{2}v^* \\
		0 & 0 & \frac{1}{2} v & \frac{1}{2} v^* & q & q & q \\
		0 & 0 & \frac{1}{2} v & \frac{1}{2} v^* & q & q & q \\
		0 & 0 & \frac{1}{2} v & \frac{1}{2} v^* & q & q & q
		\end{bmatrix}
\eeq

\subsection{Nonradiative Decay / Excitation}

Nonradiative decay is mediated through a term of the form $L = \sqrt{\gamma_{nr}}\,c_k,\ \sqrt{\bar{\gamma}_{nr}}\,\bar{c}_k$.  Strictly speaking, one must include write $L = \sqrt{\gamma_{nr}} c_k r_l^\dagger$, etc.\ where $r_l$ is the electronic mode into which the carrier decays, to make the $L$ operator bosonic.  However, if there are many more recombination sites than carriers, this mode's dynamics are not relevant and the fermionic $L$ gives the right result.

\beq
	C^{(1)} = \begin{bmatrix}
		0 \\ 0 \\ \frac{1}{2}(\gamma_{nr}+\bar{\gamma}_{nr})v \\
		\frac{1}{2}(\gamma_{nr}+\bar{\gamma}_{nr})v \\ -\gamma_{nr}m \\ -\bar{\gamma}_{nr}\bar{m} \\
		-(\gamma_{nr}+\bar{\gamma}_{nr})q
		\end{bmatrix},\ \ \
	C^{(2)} = \begin{bmatrix}
		0 & 0 & 0 & 0 & 0 & 0 & 0 \\
		0 & 0 & 0 & 0 & 0 & 0 & 0 \\
		0 & 0 & 0 & \!\!\!\!\begin{array}{c} \frac{1}{2}\gamma_{nr}(N-\bar{m}) \\ +\ \frac{1}{2}\bar{\gamma}_{nr}(N-m) \end{array}\!\!\!\! & \frac{1}{2}\gamma_{nr} v & \frac{1}{2}\bar{\gamma}_{nr} v & \frac{1}{2}(\gamma_{nr}+\bar{\gamma}_{nr})v \\
		0 & 0 & \!\!\!\!\begin{array}{c} \frac{1}{2}\gamma_{nr}(N-\bar{m}) \\ +\ \frac{1}{2}\bar{\gamma}_{nr}(N-m) \end{array}\!\!\!\! & 0 & \frac{1}{2}\gamma_{nr} v^* & \frac{1}{2}\bar{\gamma}_{nr} v^* & \frac{1}{2}(\gamma_{nr}+\bar{\gamma}_{nr})v^* \\
		0 & 0 & \frac{1}{2}\gamma_{nr} v & \frac{1}{2}\gamma_{nr} v^* & \gamma_{nr}m & 0 & \gamma_{nr}q \\
		0 & 0 & \frac{1}{2}\bar{\gamma}_{nr} v & \frac{1}{2}\bar{\gamma}_{nr} v^* & 0 & \bar{\gamma}_{nr}\bar{m} & \bar{\gamma}_{nr}q \\
		0 & 0 & \frac{1}{2}(\gamma_{nr}+\bar{\gamma}_{nr})v & \frac{1}{2}(\gamma_{nr}+\bar{\gamma}_{nr})v^* & \gamma_{nr}q & \bar{\gamma}_{nr}q & (\gamma_{nr}+\bar{\gamma}_{nr})q
		\end{bmatrix}
\eeq

\subsection{Scattering}

The scattering terms are $L = \sqrt{\gamma_{sc}/2N}\,c_\k^\dagger c_\l,\ \sqrt{\bar{\gamma}_{sc}/2N}\, \bar{c}_\k^\dagger \bar{c}_\l$.  Defining an average scattering rate by $\frac{1}{2}(\gamma_{sc} + \bar{\gamma}_{sc}) \rightarrow \gamma_{sc}$, we have:

\beq
	C^{(1)} = \gamma_{sc}\begin{bmatrix} 0 \\ 0 \\ -\frac{1}{2} v \\ -\frac{1}{2} v^* \\ 0 \\ 0 \\
		\frac{m\bar{m}}{N}-q \end{bmatrix},\ \ \
	C^{(2)} = \gamma_{sc}\begin{bmatrix}
			0 & 0 & 0 & 0 & 0 & 0 & 0 \\
			0 & 0 & 0 & 0 & 0 & 0 & 0 \\
			0 & 0 & \frac{1}{N} v^2 & \frac{1}{2}N\,e_1 & 0 & 0 & \frac{1}{2} v\,e_2 \\
			0 & 0 & \frac{1}{2}N\,e_1 & \frac{1}{N} v^2 & 0 & 0 & \frac{1}{2} v^*e_2 \\
			0 & 0 & 0 & 0 & 0 & 0 & 0 \\
			0 & 0 & 0 & 0 & 0 & 0 & 0 \\
			0 & 0 & \frac{1}{2} v\,e_2 & \frac{1}{2} v^*\,e_2 & 0 & 0 & q + \frac{m\bar{m}}{N} + e_3
		\end{bmatrix} \label{eq:app2}
\eeq
where $e_1 = 1-(m+\bar{m})/N+2m\bar{m}/N^2$, $e_2 = 1 + (2q-m-\bar{m}-2/3)/N$, and $e_3 = q(q-2(m+\bar{m}))/N$.  In Section \ref{sec:approx}, we take the limit $m, \bar{m}, q \ll N$.  In this limit, $e_1 = e_2 = 1$, $e_3 = 0$.

\newpage

\end{widetext}

\section{Related Models}
\label{sec:related}

Equations (\ref{eq:05-csde-1}-\ref{eq:05-csde-4}) are the simplest free-carrier model: identical modes, no two-photon absorption, no interaction between carriers, no excitons.  In many ways it is unrealistic.  However, it forms the basis for generalized models that include these effects and better approximate the real system.

\subsection{Non-Identical Modes}
\label{sec:manymodes}

The most obvious generalization is to include many non-identical carrier modes.  This means that, rather than grouping all of the modes together into $(m, \bar{m})$, they are binned into spectrum of modes $(m_\a, \bar{m}_\a)$.  A similar binning technique is used in many-atom cavity QED when the atomic couplings are not equal \cite{Kwon2013}.  The equations are a straightforward generalization of (\ref{eq:05-csde-1}-\ref{eq:05-csde-4}):

\bea
	d\alpha & = & \Bigl[-\frac{\kappa+\eta}{2} - i\Delta_c - i\sum_\a(\delta_\a m_\a+\bar{\delta}_\a \bar{m}_\a)\Bigr]\alpha\,dt - d\xi_\alpha \nonumber \\
	& & \\
	dm_\a & = & \Bigl[\eta_\a\,\alpha^*\alpha - \gamma_{nr,\a}m_\a - \gamma_{rc,\a}m_\a \bar{m}_\a \nonumber \\
	& & + \sum_\b{(\gamma_{\b\rightarrow\a} m_\b - \gamma_{\a\rightarrow\b} m_\a)} \Bigr]dt + d\xi_{m,\a} \\
	d\bar{m}_\a & = & \Bigl[\eta_\a\,\alpha^*\alpha - \bar{\gamma}_{nr,\a}\bar{m}_\a - \gamma_{rc,\a}m_\a \bar{m}_\a \nonumber \\
	& & + \sum_\b{(\bar{\gamma}_{\b\rightarrow\a} \bar{m}_\b - \bar{\gamma}_{\a\rightarrow\b} \bar{m}_\a)}\Bigr]dt + d\xi_{\bar{m},\a} \label{eq:05-db2}
\eea

The only change here is the introduction of indices and the cross-scattering terms $\gamma_{\a\rightarrow\b}$.  These terms, like the other carrier excitation / decay terms, have Poisson statistics.  The Poisson statistics of different modes are, of course, correlated just as the flows are -- this conserves total carrier number in the scattering processes.

If scattering between bins $m_\a$ is fast compared to carrier excitation or decay, we can replace the mode occupations by the thermal average $m_\a = f_{e,\a}(T) m, \bar{m}_\a = f_{h,\a}(T) \bar{m}$, where $f_{e,\a}, f_{h,\a}$ are normalized Boltzmann distributions.  In terms of the total carrier numbers $m = \sum_\a m_\a$ and $\bar{m} = \sum_\a \bar{m}_\a$, we recover Equations (\ref{eq:05-csde-1}-\ref{eq:05-csde-4}), with the effective rates:

\begin{align}
	\delta & = \sum_\a \delta_\a f_{\a}(T) & \bar{\delta} & = \sum_\a \delta_\a \bar{f}_{\a}(T) \nonumber \\
	\gamma_{nr} & = \sum_\a \gamma_{nr,\a} f_{\a}(T) & \bar{\gamma}_{nr} & = \sum_\a \bar{\gamma}_{nr,\a} \bar{f}_{\a}(T) \nonumber \\
	\gamma_{rc} & = \sum_\a \gamma_{rc,\a} f_\a(T) \bar{f}_\a(T)
\end{align}

\subsection{Other Processes: Kerr, TPA, FCA}
\label{sec:otherprocess}

A host of additional processes may be relevant in semiconductor cavities: among the most important are the Kerr effect, two-photon absorption (TPA), and free-carrier absorption (FCA).  Thermal effects and excitonic effects, while very important for some systems, are beyond the scope of this paper.

\subsubsection{TPA and Kerr}

In indirect-gap materials, like silicon, the linear absorption is not an effective pathway for carrier generation.  Instead, two photon absorption is the dominant excitation process.  Typically, two-photon absorption also comes with a dispersive (Kerr) effect.  In other cases, the band gap is tuned to be very close to the photon energy, and both processes are important.  Unlike linear absorption, which tends to create carriers very close to the band gap, two-photon absorption tends to create highly excited carriers with excess kinetic energy.  After excitation, these carrier quickly thermalize and subsequently decay.

We can model this with the following Hamiltonian and decay process:

\bea
	H & = & \frac{1}{2}\Delta_\x (n_\x + \bar{n}_\x) + i g \sum_\x \bigl((a^\dagger)^2 \sigma_{-\x} - a^2 \sigma_{+\x}\bigr) \nonumber \\
	\\
	L_{\x\rightarrow\k} & = & \sqrt{\gamma_{th}} c_\x c_\k^\dagger,\ \ \sqrt{\bar{\gamma}_{th}} \bar{c}_\x \bar{c}_\k^\dagger
\eea
where the new modes $c_\x, \bar{c}_\x$ defined for the highly excited carriers.  Note that, as these modes are highly excited, there is no process $L_{\k\rightarrow\x}$.

Since the excited state is so short-lived, it can be adiabatically eliminated.  For on-resonant transitions $\Delta_\x = 0$ this gives a two-photon absorption term $\beta$; in the off-resonant case $\Delta_\x \neq 0$, one finds two photon absorption plus a dispersive $\chi^{(3)}$ (Kerr) term.

These effects add the following terms to the Wigner equations:

\bea
	\!\!\!\Delta(d\alpha) & = & (-i\chi - \beta) (\alpha^*\alpha)\alpha\,dt - 2\sqrt{\beta}\alpha^* d\beta_\beta \label{eq:tpk1} \\
	\!\!\!\Delta(dm) & = & \beta(\alpha^*\alpha)^2 dt + \sqrt{\beta}\left((\alpha^*)^2d\beta_\beta + \alpha^2 d\beta_\beta^*\right) \label{eq:tpk2} \\
	\!\!\!\Delta(d\bar{m}) & = & \beta(\alpha^*\alpha)^2 dt + \sqrt{\beta}\left((\alpha^*)^2d\beta_\beta + \alpha^2 d\beta_\beta^*\right) \label{eq:tpk3}
\eea

Note how Equations (\ref{eq:tpk1}-\ref{eq:tpk3}) predict that single electron-hole pair is created for every {\it two} photons absorbed, in contrast to linear absorption (\ref{eq:05-csde-1}-\ref{eq:05-csde-4}), where the ratio is one-to-one.

\subsubsection{Free-Carrier Absorption}

In some materials, including silicon, free carriers can increase the absorption of the medium, an effect known as free-carrier absorption.  In addition, for indirect band-gap materials, the free-carrier dispersion is larger than the band-filling result (\ref{eq:delta-bf}) predicts, due to the collective response of the free-carrier plasma \cite{Bennett1990}.  These effects can be accounted for by adding the phenomenological terms:

\bea
	H & = & -i(\delta_{\rm fcd} n + \bar{\delta}_{\rm fcd} \bar{n}) a^\dagger a \\
	L & = & \sqrt{\gamma_{\rm fca} n}\, a,\ \ \sqrt{\bar{\gamma}_{\rm fca} \bar{n}}\, a
\eea

This can be accommodated in the model (\ref{eq:05-csde-1}-\ref{eq:05-csde-4}) if the substitution $\delta_c \rightarrow \delta_c + \delta_{\rm fcd} - i\gamma_{\rm fca}/2$ is made and an extra noise is included:

\beq
	\Delta(d\alpha) = -\sqrt{\gamma_{\rm fca} m + \bar{\gamma}_{\rm fca} \bar{m}}\,d\beta_{\rm fca}
\eeq
where $d\beta_{\rm fca}$ is another vacuum Wiener process.

Altogether, the Wigner equations for the free-carrier cavity, including $\chi^{(3)}$, two-photon absorption, FCD and FCA, take the form:

\begin{widetext}

\begin{align}
	d\alpha & = \left[-\frac{\kappa}{2} - i(\Delta_c + \delta_c\,m+\bar{\delta}_c\,\bar{m})\right]\alpha\,dt + \underbrace{\left[-\frac{\eta}{2}\alpha\,dt - \sqrt{\kappa} d\beta_\eta\right]}_{(d\alpha)_\eta} + \underbrace{\left[(-i\chi-\beta)(\alpha^*\alpha)\alpha\,dt - 2\sqrt{\beta}\,\alpha^*d\beta_\beta\right]}_{(d\alpha)_\beta} - \sqrt{\gamma_{\rm fca} m + \bar{\gamma}_{\rm fca} \bar{m}}\,d\beta_{\rm fca} \label{eq:sde-full1} \\
	dm & = \bigl[-\gamma_{nr}m\,dt - \sqrt{\gamma_{nr} m}\,dw_{m}\bigr] + \bigl[-\gamma_{rc}m\bar{m}\,dt + \sqrt{\gamma_{rc}m\bar{m}}\,dw_{rc}\bigr]
		- \left[\alpha^*(d\alpha)_\eta + \alpha(d\alpha)_\eta^*\right]
		- \frac{\alpha^*(d\alpha)_\beta + \alpha(d\alpha)_\beta^*}{2} \label{eq:sde-full2} \\
	d\bar{m} & = \bigl[-\bar{\gamma}_{nr}\bar{m}\,dt - \sqrt{\bar{\gamma}_{nr} \bar{m}}\,dw_{\bar{m}}\bigr] + \bigl[-\gamma_{rc}m\bar{m}\,dt + \sqrt{\gamma_{rc}m\bar{m}}\,dw_{rc}\bigr]
		- \left[\alpha^*(d\alpha)_\eta + \alpha(d\alpha)_\eta^*\right]
		- \frac{\alpha^*(d\alpha)_\beta + \alpha(d\alpha)_\beta^*}{2} \label{eq:sde-full3}
\end{align}

\subsection{Single-Carrier Approximation}

The single-carrier approximation assumes $m = \bar{m} = N$, and replaces $\delta_c + \bar{\delta_c} \rightarrow \delta_c$, $\gamma_{\rm fca} + \bar{\gamma}_{\rm fca} \rightarrow \gamma_{\rm fca}$.  The equations reduce to:

\begin{align}
	d\alpha & = \left[-\frac{\kappa}{2} - i(\Delta_c + \delta_c N_c) \right]\alpha\,dt + \underbrace{\left[-\frac{\eta}{2}\alpha\,dt - \sqrt{\kappa} d\beta_\eta\right]}_{(d\alpha)_\eta} + \underbrace{\left[(-i\chi-\beta)(\alpha^*\alpha)\alpha\,dt - 2\sqrt{\beta}\,\alpha^*d\beta_\beta\right]}_{(d\alpha)_\beta} - \sqrt{\gamma_{\rm fca} N_c}\,d\beta_{\rm fca} \label{eq:sde-sc1-b} \\
	dN_c & = \bigl[-\gamma_{nr}N_c\,dt - \sqrt{\gamma_{nr} N_c}\,dw_{nr}\bigr] + \bigl[-\gamma_{rc}N_c^2\,dt + \sqrt{\gamma_{rc}N_c^2}\,dw_{rc}\bigr]
		- \left[\alpha^*(d\alpha)_\eta + \alpha(d\alpha)_\eta^*\right]
		- \frac{\alpha^*(d\alpha)_\beta + \alpha(d\alpha)_\beta^*}{2} \label{eq:sde-sc2-b}
\end{align}

\end{widetext}

\section{Carrier Detuning in terms of Material Properties}
\label{sec:materials}

In this section we derive expressions for the coupling constant $g_\k$ and the carrier-dependent detuning $\delta_\k$ as a function of material properties.  This is important because it allows one to match the results from this work to the semiclassical treatment of FCD found elsewhere in the literature.

In standard single-particle electrodynamics, to first order in the optical field the light-matter coupling goes as:

\beq
	H_{\rm int} = \frac{e \vec{A} \cdot \vec{p}}{m_0} \label{eq:eap}
\eeq
This can be generalized to many-particle systems by ``second-quantizing'' the Hamiltonian in terms of fermionic creation / annihilation operators $f_\k, f_\k^\dagger$ \cite{KiraKoch2012}:

\beq
	H_{\rm int} \rightarrow \frac{e}{m_0} \sum_{\k,\l\in\rm \{states\}}{f_\k^\dagger \bracket{\k}{A \cdot p}{\l} f_\l} \label{eq:05-hint}
\eeq

Consider a two-band model.  The $f_\k$ here represent both valence-band and conduction-band states.  If the field $A$ is driving at optical frequencies, only transitions between the valence band and conduction band need be considered -- for these, the Hamiltonian becomes:

\bea
	H_{\rm int} & \rightarrow & \frac{e}{m_0} \sum_{\k} \Bigl[c_{\k}^\dagger \bar{c}_{\k}^\dagger \bbracket{\k,c}{\vec{A}(x,t) \cdot \vec{p}}{\k,v} \nonumber \\
	& & \qquad\qquad + \bar{c}_{\k} c_{\k} \bbracket{\k,v}{\vec{A}(x,t) \cdot \vec{p}}{\k,c}\Bigr] \label{eq:05-hint2}
\eea

For a resonant structure, $E(x, t)$ and $B(x, t)$ depend on the normal-mode fields $E_\omega(x)$ and their time-dependent amplitude $a_\omega(t)$ (which becomes the photon annihilation operator when the system is quantized).  Working in the Coulomb gauge $\phi(x,t) = 0$, $E = -\partial A/\partial t$ and $A(x, t)$ is given by:

\beq
	\vec{A}(x, t) = \mbox{Re}\left[\sum_\omega -i \sqrt{2\hbar/\omega\epsilon_0} a_\omega \vec{E}_\omega(x) e^{-i\omega t}\right] \eeq
	
Here $E_\omega$ is normalized so that $\int{n(x)^2 |E_\omega|^2 d^3x} = 1$, and $a_\omega$ is the photon annihilation operator.  For a good resonator, typically only one frequency $\omega$ is relevant (though multiple frequencies is a simple extension of this work), so hereafter we replace $a_\omega \rightarrow a$.  Going into the interaction picture and neglecting rotating-wave terms and adding an arbitrary phase shift to the $c_i$ to fix the sign of $g_\k$, we find:

\beq
	H_{\rm int} = \sum_{\k} ig_{\k} \left(a^\dagger c_{\k}\bar{c}_{\k} - a\,\bar{c}_{\k}^\dagger c_{\k}^\dagger\right)
\eeq
with coupling constant $g_{\k}$ given by:

\beq
	g_{\k} = \frac{e}{m_0} \sqrt{\frac{\hbar}{2\omega\epsilon_0}}\ \Bigl| \vec{E}_\omega(x) \cdot \sbracket{\k,c}{\vec{p}}{\k,v} \Bigr|
\eeq

The electronic and photon parts to the Hamiltonian take their canonical forms.  The end result is (\ref{eq:05-ham}).

Having derived the coupling $g_\k$, we proceed to express the carrier-dependent detuning in (\ref{eq:delta-bf}) in terms of actual material properties.  The carrier-dependent detuning is what fundamentally limits the performance of a free-carrier device -- it sets the minimum number of carriers needed to switch by one linewidth, the energy figure of merit for a photonic switch.  It is given by:

\beq
	\Delta(m,\bar{m}) = \sum_\k(\delta_\k m_\k + \bar{\delta}_\k \bar{m}_\k)
\eeq
with $\delta_\k = g^2/(\Delta_\k-\frac{1}{2}i\gamma_{sc})$, as in (\ref{eq:delta-bf}).

This section considers two common cases: a III-V semiconductor near the band gap, where band filling is dominant, and silicon far from the band gap, where the plasma effect dominates.  These effects are well studied in bulk materials; the point of this section is to translate them to the optical resonator picture used in this paper.

\subsection{III-V Semiconductor near Band Gap}

Here, the dominant effect comes from band-filling dispersion.  We assume that all modes have roughly the same energy, $E_g$, and that the optical field is at $E = x E_g$, where $x < 1$.  If $x \approx 1$, then one can show that the carrier-dependent detuning takes the form:

\bea
	\Delta(m,\bar{m}) & \approx & \sum_\k{\delta_{\k}(m_\k+\bar{m}_\k)} \nonumber \\
	& \approx & \sum_\k\frac{\hbar^2 e^2 |\vec{E}_\omega(x_\k) \cdot \vec{p}_{cv}|^2}{m_0^2 \epsilon_0 E_g^2} \frac{1}{2x(1-x)}(m_\k + \bar{m}_\k) \nonumber \\
	& = & \frac{\hbar^2 e^2 |E_\omega(x_\k)|^2 |p_{cv}|^2}{m_0^2 \epsilon_0 E_g^2} |\hat{E}_\omega(x_\k)\cdot\hat{p}_{cv}|^2 \nonumber \\
	& & \quad \times \frac{1}{2x(1-x)} (m_\k + \bar{m}_\k)
\eea
where $m_0$ is the bare electron mass and $\vec{p}_{cv}$ is the matrix element $\bra{\k,c} \vec{p} \ket{\k, v}$ between conduction- and valence-band states

\begin{figure}[tbp]
\begin{center}
\includegraphics[width=0.8\columnwidth]{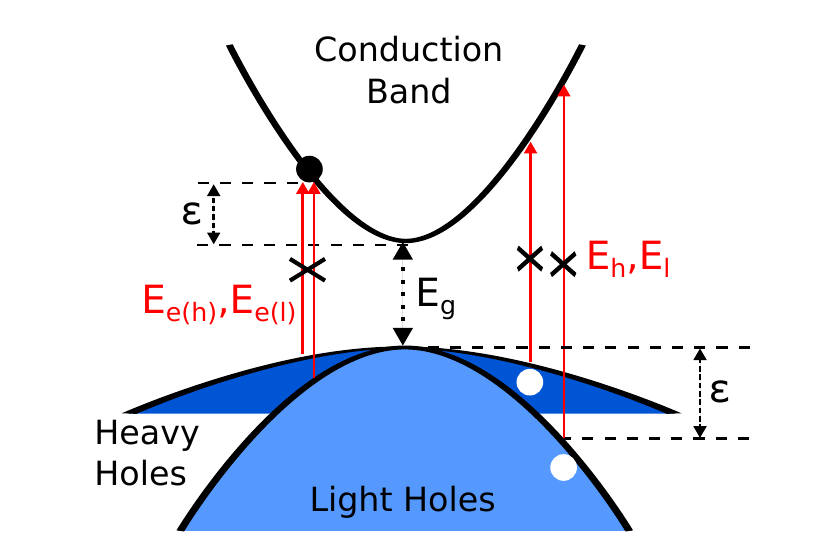}
\caption{Bandfilling in a direct-gap III-V semiconductor.  Carriers block certain optical transitions, changing the absorption spectrum, which in turn alters the index of refraction.}
\label{fig:01a-f1}
\end{center}
\end{figure}

This is a two-band calculation, which only includes transitions from a single valence band.  Adding a second valence band doubles the effect of the electrons -- since each electron ``blocks'' two transitions, one from each valence band, its bandfilling effect is doubled (Figure \ref{fig:01a-f1}).  This does not happen for holes, since each hole only ``blocks'' the one transition to the conduction band.  Thus the correct carrier-dependent detuning is:

\bea
	\Delta(m,\bar{m}) & \approx & \sum_\k \frac{3\hbar^2 e^2 |E_\omega(x_\k)|^2 |p_{cv}|^2}{m_0^2 \epsilon_0 E_g^2} |\hat{E}_\omega(x_\k)\cdot\hat{p}_{cv}|^2 \nonumber \\
	& & \quad \times \frac{1}{2x(1-x)} \frac{2m_\k + \bar{m}_\k}{3}
\eea

To get a sense of scaling, we replace $|E_\omega|^2 \rightarrow |\tilde{E}_\omega|^2/(n_0^2 V)$.  Here, $\tilde{E}_\omega$ is designed to have near-unit amplitude within the cavity, and $V$ is the mode volume.  Unlike $E_\omega$, $\tilde{E}_\omega$ is not normalized (its integral is not one), but having near unit-amplitude is what matters here.  The mode volume is defined in terms of a normalized quantity, $\tilde{V} = V/(\lambda/n)^3$, which is $O(1)$ for photonic crystals and $O(10)$ for rings.  Instead of looking at $\Delta$, we look at $\Delta/\omega$, since this is unitless, and we are well aware that $\Delta(m,\bar{m})/\omega \sim 1/Q$ means that enough carriers have been injected to move the cavity one linewidth.

\beq
	\frac{\Delta(m,\bar{m})}{\omega} = \frac{3e^2 n_0 |p_{cv}|^2}{8\pi^3 m_0^2 \hbar c^3 \epsilon_0 \tilde{V}} \frac{x}{2(1-x)} \sum_\k{|\tilde{E}(x_\k)\cdot\hat{p}_{cv}|^2 \frac{2m_\k + \bar{m}_\k}{3}} \label{eq:05-del}
\eeq

This is a product of four terms.  (1) The first is a bunch of fundamental constants, plus material and cavity parameters like the cavity size and the index of refraction, and the magnitude of the matrix element $|p_{cv}|$.  These constants depend only on the device, not on the wavelength operated at or the particular carrier mode being excited.  (2) Next we have a term that depends on the closeness to the band edge: $x/2(1-x)$.  In practice, we will want $x$ to be as close to one as possible in order to maximize this quantity -- however, as $x \rightarrow 1$ linear absorption limits the cavity $Q$, so there is a tradeoff here.  (3) Next is a position term that depends on the field strength at $x_\k$, the location of the carrier (assuming carriers localized to well below a wavelength here).  (4) Finally, the carrier number.

When carrier thermalization and diffusion is fast compared to the decay processes, one can write this as an effective single-mode model, where the per-carrier detuning is given by the thermal average:

\begin{align}
	& \frac{\Delta(m,\bar{m})}{\omega} \nonumber \\
	& \ \ \rightarrow \frac{3e^2 n_0 |p_{cv}|^2}{8\pi^3 m_0^2 \hbar c^3 \epsilon_0 \tilde{V}} \frac{x}{2(1-x)} \avg{|\tilde{E}(x_\k)\cdot\hat{p}_{cv}|^2}_\k \frac{2m + \bar{m}}{3} \nonumber \\
	& \ \ = \frac{e^2 n_0 |p_{cv}|^2}{8\pi^3 m_0^2 \hbar c^3 \epsilon_0 \tilde{V}} \frac{x}{2(1-x)} \langle|\tilde{E}(x_\k)|^2\rangle_\k \frac{2m + \bar{m}}{3} \label{eq:05-bfm}
\end{align}

In the limit $x \approx 1$, this is consistent with previous derivations of the band-filling dispersion \cite{Bennett1990, Said1992}, under the replacements $|p_{cv}|^2 \rightarrow E_g m_0^2/2m_e$ and $\langle|\tilde{E}(x)|^2\rangle_\k \rightarrow 1$ (this is always $O(1)$ and the equality can be imposed by scaling $\tilde{V}$).  Because of the rotating-wave approximation taken in this paper, it will not be valid when $x$ deviates far from 1.  However, optimized devices exploiting band-filling always operate near the band gap.

\subsection{FCD in Silicon}

In silicon, the indirect band gap makes the band-filling effect very weak.  Instead, free-carrier dispersion is dominated by the plasma effect \cite{Bennett1990}.  Consider a simple Drude model with a carrier density $N$.  The index of refraction is modified as follows:

\beq
	n^2 \rightarrow n_0^2\left(1 - \frac{n_c e^2/m_c n_0^2 \epsilon_0}{\omega^2 + i\omega/\tau}\right)
\eeq

where $n_c = n, p$ are the densities and $m_c = m_e, m_h$ are the masses for electrons and holes.  In the high-frequency limit $\omega \gg \omega_p$, the real part dominates and this becomes:

\beq
	\Delta n = -\frac{\hbar^2e^2}{2n_0\epsilon_0 E^2} \left[\frac{n}{m_e} + \frac{p}{m_h}\right]
\eeq

Assuming the carriers are confined to a volume $V$, and defining the dimensionless $\tilde{V} = V/(\lambda/n)^3$ as above, and using standard coupled-mode theory to convert $\Delta n$ to a detuning, we find:

\beq
	\Delta(m,\bar{m}) = \frac{e^2 n_0}{16\pi^3 \hbar c^3 \epsilon_0 \tilde{V}} \left[\frac{m}{m_e} + \frac{\bar{m}}{m_h}\right]
\eeq

A more detailed treatment shows that the dependence is linear for electrons, but nonlinear for holes \cite{Soref1987}.  This nonlinearity can be treated phenomenologically in (\ref{eq:05-csde-1}-\ref{eq:05-csde-4}); the quantum noise terms derived in this section do not change.

\end{document}